\documentclass[sigconf,nonacm]{acmart}
\AtBeginDocument{%
  \providecommand\BibTeX{{%
    \normalfont B\kern-0.5em{\scshape i\kern-0.25em b}\kern-0.8em\TeX}}}

\acmSubmissionID{68}

\usepackage{algorithmic}
\usepackage{graphicx}
\usepackage{textcomp}
\usepackage{xcolor}
\usepackage{comment}
\usepackage{url}
\usepackage{caption}
\usepackage[ruled,linesnumbered]{algorithm2e} 
\SetAlCapNameFnt{\small}
\SetAlCapFnt{\small}
\usepackage{siunitx}
\usepackage{graphicx}
\usepackage{caption}
\usepackage{subcaption}
\begin{document}


\title{CBP: Coordinated management of cache partitioning, bandwidth partitioning and prefetch throttling
}


   \author{Nadja Ramhöj Holtryd, Madhavan Manivannan, Per Stenstr{\"o}m, Miquel Peric{\`a}s} 
    \affiliation{ 
      \institution{Department of Computer Science and Engineering}
      \streetaddress{street}
      \city{Chalmers University of Technology} \\
      \state{G{\"o}teborg} 
      \postcode{5 positive integers}
      \country{Sweden}
    }
    \email{Email: { holtryd, madhavan, per.stenstrom, miquelp } @chalmers.se}


\begin{abstract}
Reducing the average memory access time is crucial for improving the performance of applications running on multi-core architectures. With workload consolidation this becomes increasingly challenging due to shared resource contention. Techniques for partitioning of shared resources - cache and bandwidth - and prefetching throttling have been proposed to mitigate contention and reduce the average memory access time. However, existing proposals only employ a single or a subset of these techniques and are therefore not able to exploit the full potential of coordinated management of cache, bandwidth and prefetching. 
Our characterization results show that application performance, in several cases, is sensitive to prefetching, cache and bandwidth allocation. Furthermore, the results show that managing these together provides higher performance potential during workload consolidation as it enables more resource trade-offs. In this paper, we propose CBP a coordination mechanism for dynamically managing prefetching throttling, cache and bandwidth partitioning, in order to reduce average memory access time and improve performance. CBP works by employing individual resource managers to determine the appropriate setting for each resource and a coordinating mechanism in order to enable inter-resource trade-offs. Our evaluation on a 16-core CMP shows that CBP, on average, improves performance by 11\% compared to the state-of-the-art technique that manages cache partitioning and prefetching and by 50\% compared to the baseline without cache partitioning, bandwidth partitioning and prefetch throttling. 
\end{abstract}

\maketitle

\section{Introduction}
\label{intro}

Memory access time has a significant impact on application performance. 
Effective utilization of the memory system is therefore necessary. Typically, resources in the memory system (e.g., last-level cache (LLC) and off-chip memory bandwidth) are shared among multiple cores as they help in improving resource utilization during workload consolidation. However, sharing can detrimentally impact average memory access time and performance due to resource contention. Prior works have proposed partitioning of shared resources -- cache ~\cite{Qureshi2006,Lee2011,kpart18,Manikantan2012,Sanchez2011} and bandwidth~\cite{Park2018,pabst17,liu10} -- and prefetching~\cite{Falsafi2014} to mitigate contention, reduce or hide memory access time and improve performance. 

Recent works have proposed combining cache and bandwidth partitioning~\cite{Sahu2014,Bitirgen2008,copart19}, prefetching and cache partitioning~\cite{Xiao2019,Sun2019} and bandwidth partitioning and prefetching~\cite{Ebrahimi2011,Liu2011} to provide additional performance gains. The key insight from these papers is that coordinated management of two techniques is more advantageous than considering each in isolation because of the trade-offs that are made
possible. However, no study so far has considered combining all three techniques. The goal of this paper is to do so.

Coordinated management of cache partitioning, bandwidth partitioning and prefetch throttling provides the following advantages. Firstly, it makes it possible to address more applications and cover a broader range of workloads, as shown in our in-depth performance characterization (see Section \ref{sec:motivation}). The results show that 90\% of the applications in the SPEC CPU2006 suite have performance sensitivity (over 10\% change in IPC) to at least one of the techniques, and 70\% are also sensitive to multiple techniques. 
Secondly, managing these techniques jointly opens up the opportunity to new and improved trade-offs. 
\begin{figure}[b!]
\centering
\includegraphics[width=0.29\textwidth]{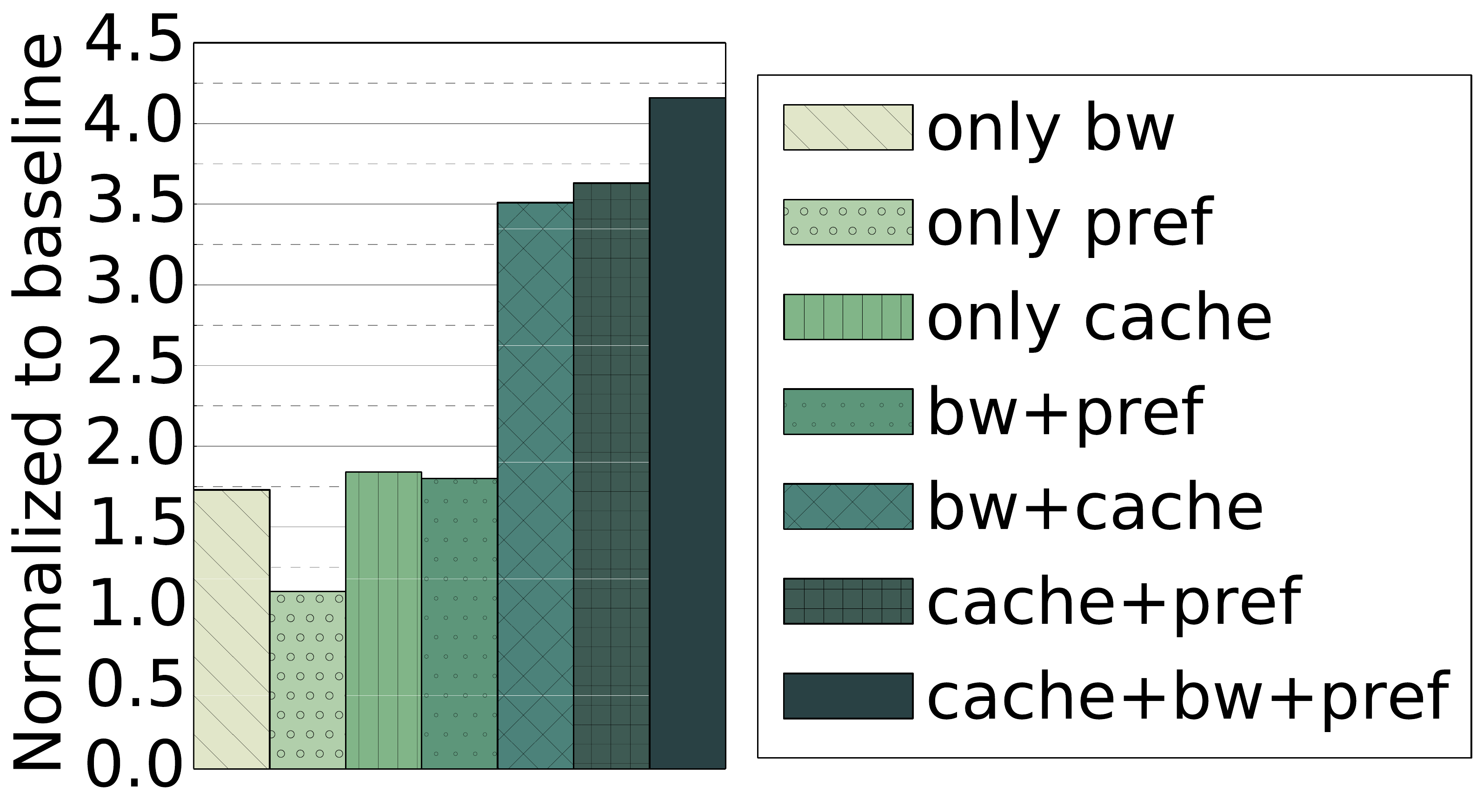}
\vspace{-1.5em}
\caption{Workload with lbm and xalancbmk. Total bandwidth 16GB/s, total cache size 2MB. Executing applications for 1B instructions, more details in Section \ref{sec:methodology}. Settings: lbm- prefething active, 12GB/s, xalancbmk- prefetcher inactive, 4GB/s, determined from characterization (Section \ref{sec:motivation}). Cache partition sizes are decided dynamically.}
\label{fig:mot_lbm_xalancbmk}
\end{figure}
There are synergistic interactions between the techniques and these cannot be realized if cache partitioning, bandwidth partitioning and prefetch throttling are not jointly managed.

As an example
, consider the simple case of a workload comprising of two applications. The first application, \textit{lbm}, is sensitive to bandwidth and prefetching while the second, \textit{xalancbmk}, is sensitive to cache size and has lower performance when prefetching is enabled. The best solution when managing all three techniques is to give xalancbmk a large cache allocation, small bandwidth allocation and disable the prefetcher, while giving lbm a large bandwidth allocation, small cache allocation while keeping the prefetcher active. Figure \ref{fig:mot_lbm_xalancbmk} shows the performance from coordinated management of all three techniques (\textit{cache+bw+pref}) compared to managing a subset of the techniques. The results show that the solution that manages all three techniques is better than others that manage two of the techniques, and leads to an additional performance gain of 15\%.
The main challenge of coordinately managing all three techniques is the complexity of evaluating all possible allocations dynamically and determining the best possible allocation while exploiting the large number of possible trade-offs. 

Guided by our characterization results, we propose CBP, a coordinated mechanism for dynamically managing cache partitioning, bandwidth partitioning and prefetch throttling for multi-programmed workloads. CBP consists of three local controllers, one for each resource that together with a coordination mechanism manages and allocates the resources. With CBP, the three techniques are dynamically tuned in an iterative fashion. 
First, cache space is allocated since avoiding a memory access altogether is better than reducing their latency. As a next step, bandwidth is allocated taking into consideration the impact due to cache allocation. Lastly, the prefetch setting is determined by testing the impact of prefetching on performance for the current allocation of bandwidth and cache. The prefetcher performance influences the next reallocation of cache space and bandwidth. The feedback mechanism between the different techniques dynamically adapts the allocations in order to reach a good configuration depending on the characteristics of the individual applications in the workload. Our approach of combining local controllers with a feedback mechanism reduces the complexity. 
 
In summary, we make the following contributions:

(a) We present an in-depth characterization of the performance impact of cache, bandwidth and prefetching on the entire SPEC CPU2006 suite. Our characterisation results provide several insights: i) a majority of the applications (over 90\%) are sensitive to one or multiple  techniques, ii) managing cache, bandwidth and prefetch, opens up opportunities for exploiting more trade-offs and improving performance for consolidated workloads, and iii) managing cache, bandwidth and prefetch jointly has the potential to outperform combinations of two of the techniques. 

(b) We propose CBP, a mechanism to dynamically manage the three resources in coordination. The solution is based on simple heuristics in order to sidestep the complexity associated with evaluating all possible configurations and choosing the most efficient configuration. CBP works by employing individual resource managers to determine the appropriate setting for each resource and a coordinating mechanism to enable inter-resource trade-offs.


(c) We evaluate our solution with multi-programmed workloads on a 16-core tiled CMP. CBP improves performance by up to 36\% (geom. mean 11\%) compared to the state-of-the-art technique that manages cache partitioning and prefetching in a coordinated manner and by up to 86\% (geom. mean 50\%) compared to an unpartitioned S-NUCA without cache partitioning, bandwidth partitioning and prefetching. 

The rest of the paper is organized as follows. Section \ref{sec:motivation} motivates the need for a coordinated approach using cache partitioning, bandwidth partitioning and prefetch throttling. 
Section \ref{sec:descriptionofproposal} describes our proposed solution in detail. We then discuss the methodology in Section \ref{sec:methodology} and  Section \ref{sec:evaluation} presents the evaluation of the proposal. We provide an overview of related work in Section \ref{sec:relatedwork} and conclude in Section \ref{sec:conclusion}.   
\vspace{-1em}
\section{Characterization} 
\label{sec:motivation}
In order to motivate the need for coordinated management of cache partitioning, bandwidth partitioning and prefetch throttling, we perform a detailed characterization study of applications in the SPEC2006 CPU suite. The aim of this study is to: i) characterize applications to determine the extent to which they are performance sensitive to cache, bandwidth and prefetch settings, ii) understand the different resource interactions, their impact on performance and the inter-resource trade-offs that are possible, and iii) demonstrate the performance potential of coordinated management of all three resources over a subset of resources.
\vspace{-1em}
\subsection{Sensitivity to cache, bandwidth and prefetch settings} 
\label{sec:sensitivity}
To understand the sensitivity of applications to cache, bandwidth and prefetch settings we model a system consisting of one out-of-order core with a  3-level cache hierarchy using the Sniper simulator \cite{Carlson:2014:EHM:2658949.2629677}. Details about the methodology are provided in Section \ref{sec:methodology}. For this experiment, the baseline LLC and bandwidth allocation is 512kB and 4GB/s, respectively. We run the application in steady-state for 1B instructions and use IPC as a measure of performance. 

\begin{figure*}[h]
\centering
\begin{subfigure}[t]{0.49\textwidth}
\centering
\includegraphics[width=\textwidth]{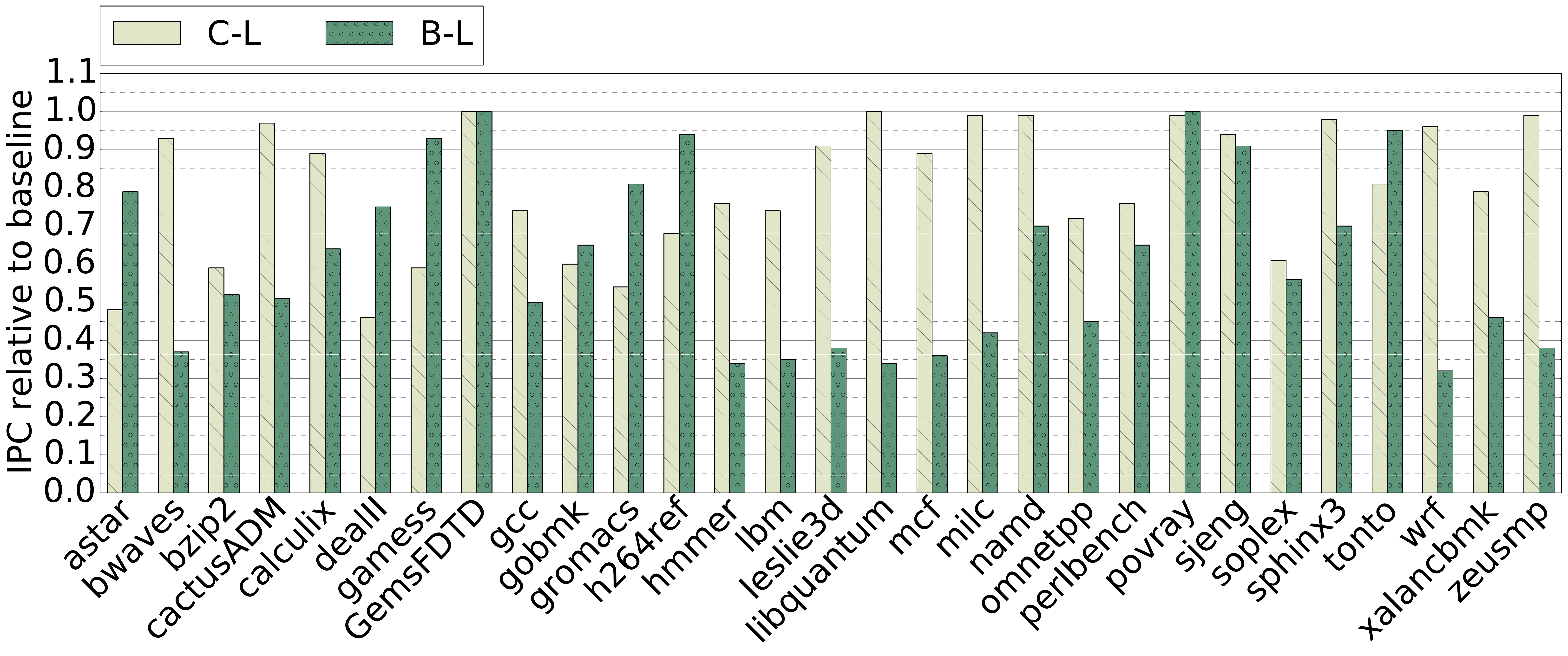}
\caption{Slowdown when decreasing the cache size to 128kB (C-L), and slowdown when decreasing the bandwidth allocation to 1GB/s (B-L) in comparison to the baseline allocation, with prefetching disabled.}
\label{fig:csbs_low}
\end{subfigure}
\hfill
\begin{subfigure}[t]{0.49\textwidth}
\centering
\includegraphics[width=\textwidth]{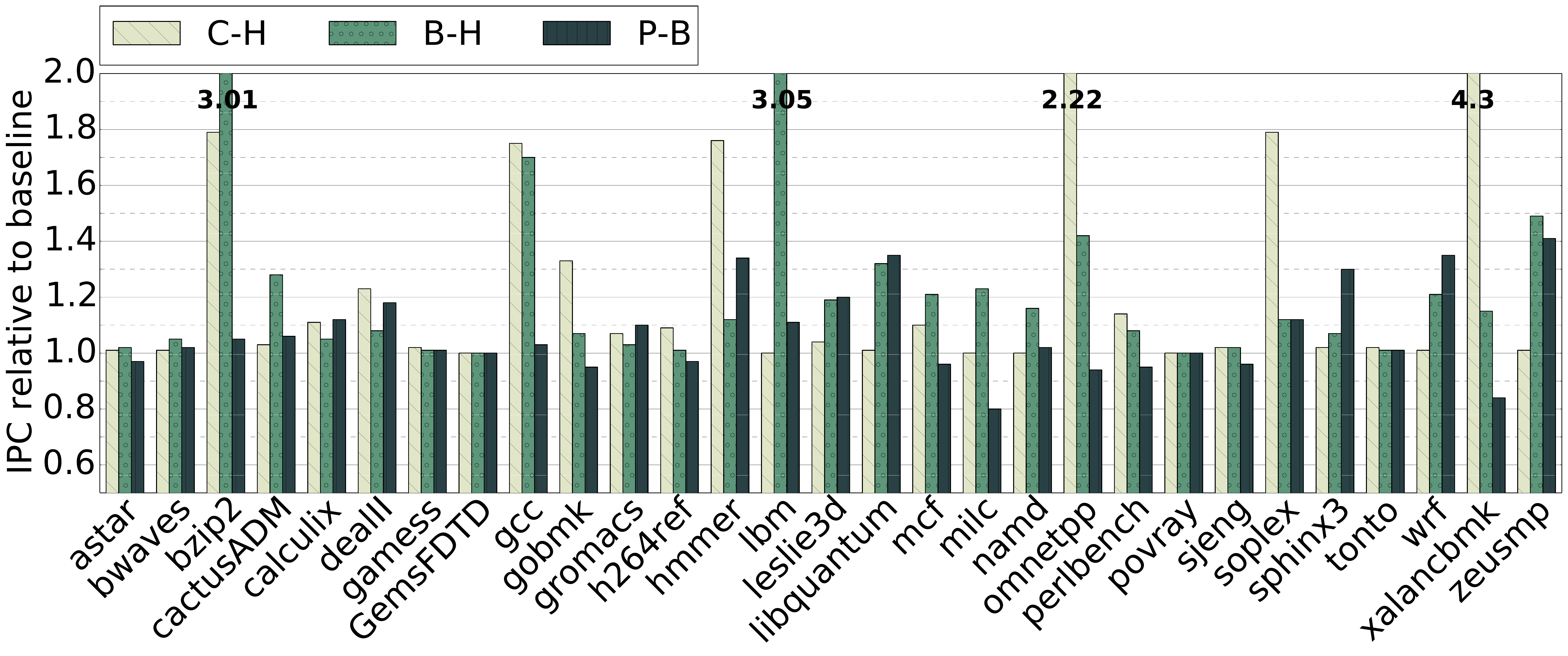}
\caption{Performance improvement from increasing cache allocation to 2MB (C-H)
, and increasing bandwidth allocation to 16GB/s (B-H) in comparison to the baseline allocation, with prefetching disabled. Performance improvement from enabling prefetching for the baseline allocation (P-B)}
\label{fig:csbsps_high}
\end{subfigure}
\caption{Performance impact of changing cache size, bandwidth allocation and prefetcher setting. There are 6 CS-BS-PS applications, 8 CS-BS, 6 BS-PS, 3 CS, 3 BS and 3 applications are insensitive (I) to all three techniques.}
\label{fig:csbs}
\end{figure*}

Figure \ref{fig:csbs} shows the performance impact of 
changing the cache allocation, bandwidth allocation and enabling prefetching normalized to the baseline allocation without prefetching. Note that we only change the setting for one resource at a time. 
In Figure \ref{fig:csbs_low}, C-L and B-L represent low allocation settings where the cache allocation is decreased to 128kB and the bandwidth allocation is decreased to 1GB/s, respectively, while prefetching is disabled. Similarly, in Figure \ref{fig:csbsps_high}, C-H and B-H represent high allocation settings where the cache allocation is increased to 2MB and the bandwidth allocation is increased to 16GB/s, while prefetching is disabled. Finally, P-B represents the setting where prefetching is enabled with the baseline cache and bandwidth allocation.
We classify applications as \textit{performance sensitive} to a specific resource if the modified allocation results in a 10\% deviation from the baseline IPC. We refer to applications that are performance sensitive to change in cache allocation as cache sensitive (CS), sensitive to change in bandwidth allocation as bandwidth sensitive (BS) and sensitive to prefetch throttling as prefetch sensitive (PS). 

The sensitivity results for cache size show that nearly 60\%  of the applications (17 out of 29) are sensitive to changes in cache allocation. The extent to which applications are performance sensitive varies greatly with a performance increase of up to 4x in some cases. Furthermore, a larger number of applications are sensitive in the low allocation setting in comparison to high allocation setting (17 compared to 11). The sensitivity results for bandwidth allocation also shows a similar trend, as more applications are sensitive in the low allocation setting (23 compared to 15). Also, the extent of performance sensitivity varies greatly with an increase of up to 3x. The sensitivity results for prefetch throttling indicate that nearly 38\% of the applications (11 out of 29) are sensitive to prefetching and experience a speedup. However, there are some applications that experience a slowdown due to prefetching. In summary, we make the following observation:
\vspace{0.2em}

\noindent \textbf{OBSERVATION 1.} \textit{In SPEC CPU2006 suite 90\% of the applications are sensitive to one resource, while 70\% are sensitive to multiple resources and the extent of sensitivity varies greatly.}

\subsection{Inter-resource interactions and trade-offs} 
\label{sec:inter}
Next, we investigate inter-resource interactions and trade-offs that are enabled when jointly managing cache, bandwidth and prefetching.
We focus on intra-application resource interaction initially. 
There are four possible types of interactions within an application: cache-bandwidth-prefetch, bandwidth - prefetch, cache - prefetch and cache - bandwidth. 

\begin{figure}[h]
\centering
\includegraphics[width=0.5\textwidth]{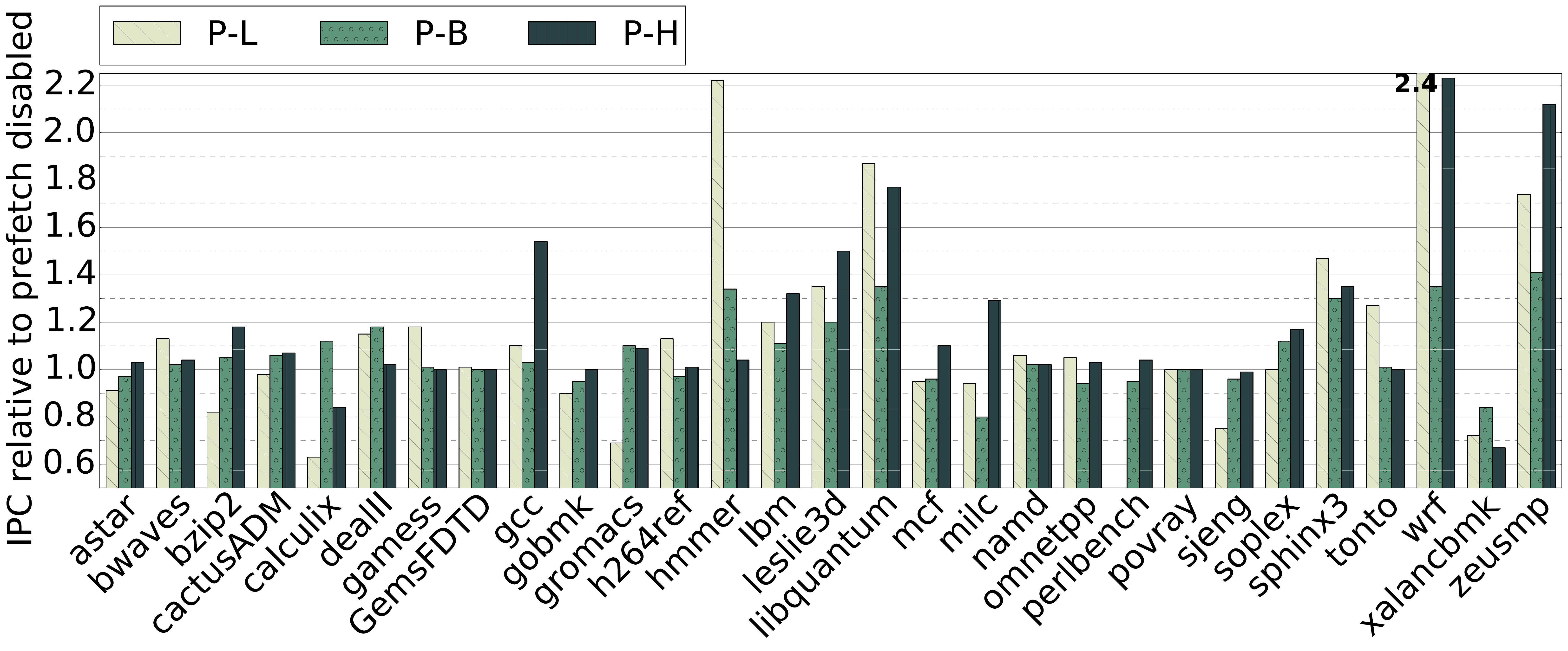}
\vspace{-2.5em}
\caption{Performance impact of enabling prefetching relative to allocation of cache and bandwidth, for allocation settings; L:128kB,1GB/s B:512kB,4GB/s H:2MB,16GB/s}
\label{fig:ps}
\end{figure}

Regarding the cache-bandwidth-prefetch trade-off, we want to find out how the performance impact from prefetching varies with the allocation of cache and bandwidth. Figure \ref{fig:ps} shows the performance impact of prefetching for three different cache/bandwidth settings normalized to the respective baseline setting without prefetching. The cache and bandwidth setting for an application in a low allocation scenario (P-L) is 128kB and 1GB/s, the baseline allocation scenario (P-B) setting is 512kB and 4GB/s while the high allocation scenario (P-H) setting is 2MB and 16GB/s. 

For some applications, lower bandwidth and cache allocation leads to higher sensitivity for prefetching as seen in \textit{hmmer}. This is because avoiding a miss altogether, as a consequence of accurate prefetching, can have a larger impact in low allocation settings where the bandwidth is scarce and the memory queuing delays tend to be 
longer. Also, there are applications like \textit{gcc} which experience higher prefetch sensitivity with a larger cache and bandwidth allocation. 
The results indicate that applications tend to be prefetch sensitive in some settings and prefetch insensitive in others. We make the following observation:
\vspace{0.2em}

\noindent \textbf{OBSERVATION 2.} \textit{ Allocation of cache and bandwidth influences prefetch sensitivity. Furthermore, applications tend to be prefetch sensitive in some settings and prefetch insensitive in others.}
\\
\vspace{-0.5em}
\begin{figure*}[h!]
\centering
\begin{subfigure}[t]{0.2\textwidth}
\centering
\includegraphics[width=1\textwidth]{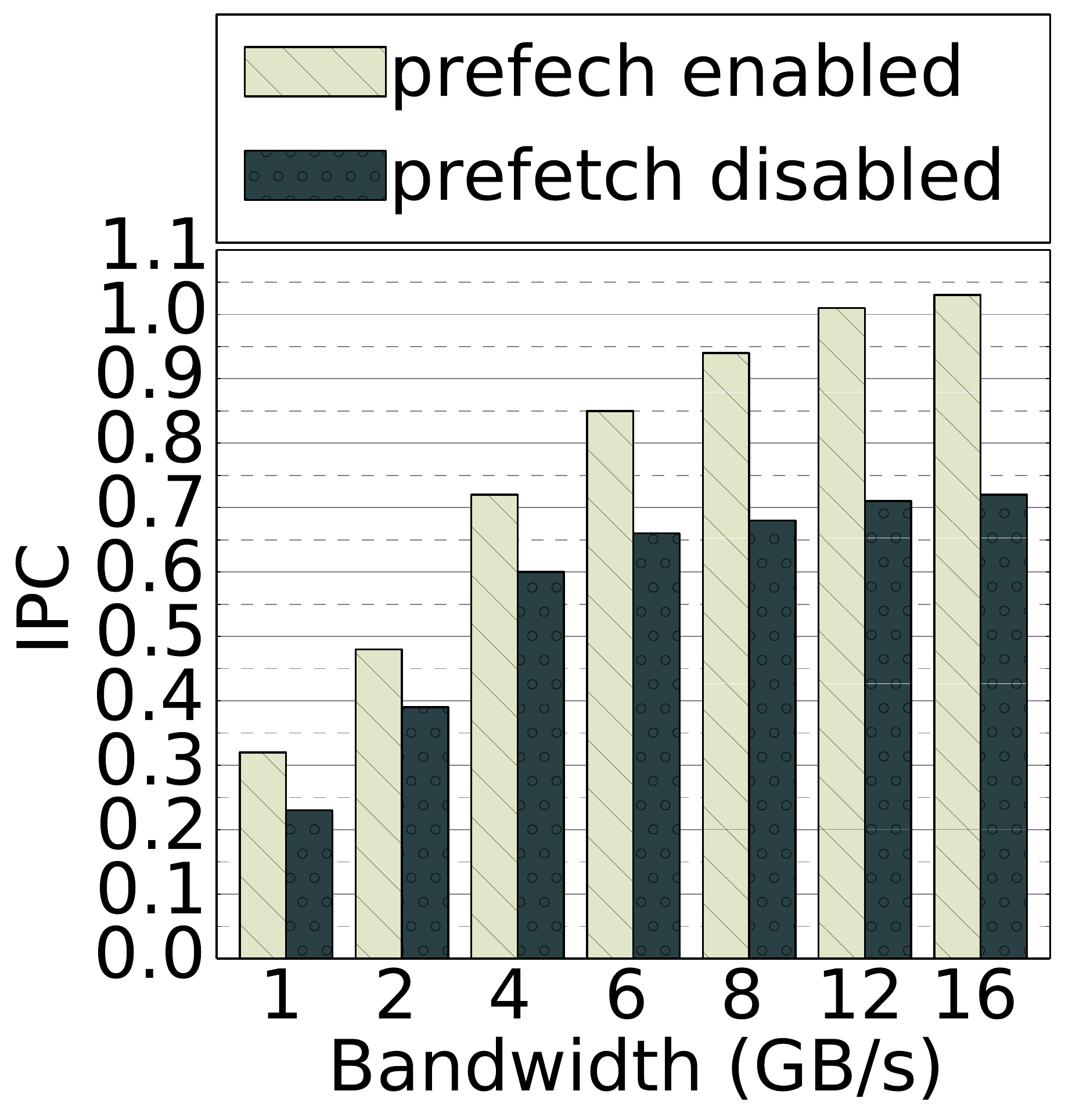}
\vspace{-0.5em}
\caption{IPC with and without prefetching for different bandwidth allocation.}
\label{fig:leslie_a}
\end{subfigure}
\hfill
\begin{subfigure}[t]{0.22\textwidth}
\centering
\includegraphics[width=1\textwidth]{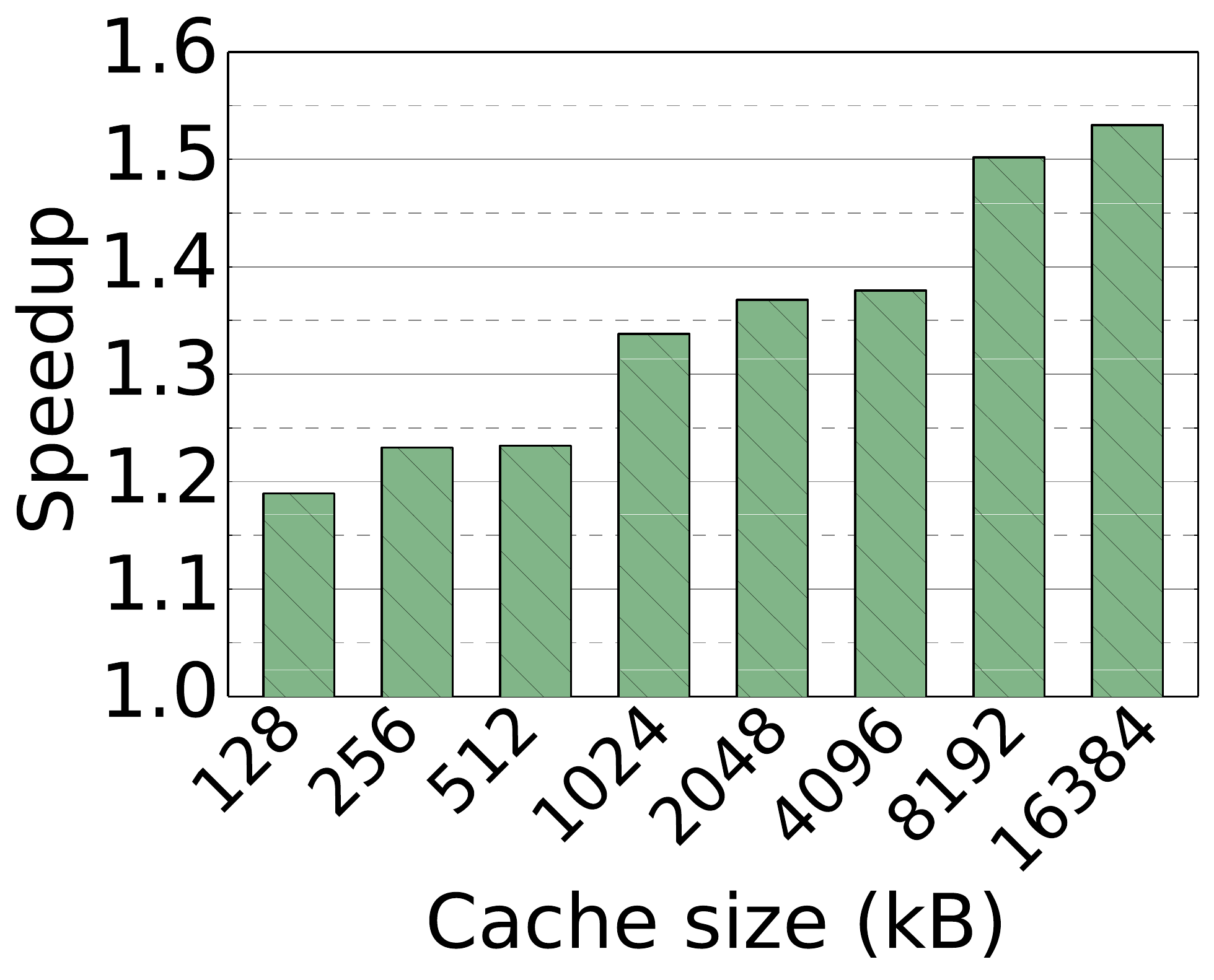}
\vspace{-0.5em}
\caption{Performance improvement from prefetching depending on cache allocation.}
\label{fig:leslie3d_b}
\end{subfigure}
\hfill
\begin{subfigure}[t]{0.23\textwidth}
\centering
\includegraphics[width=1\textwidth]{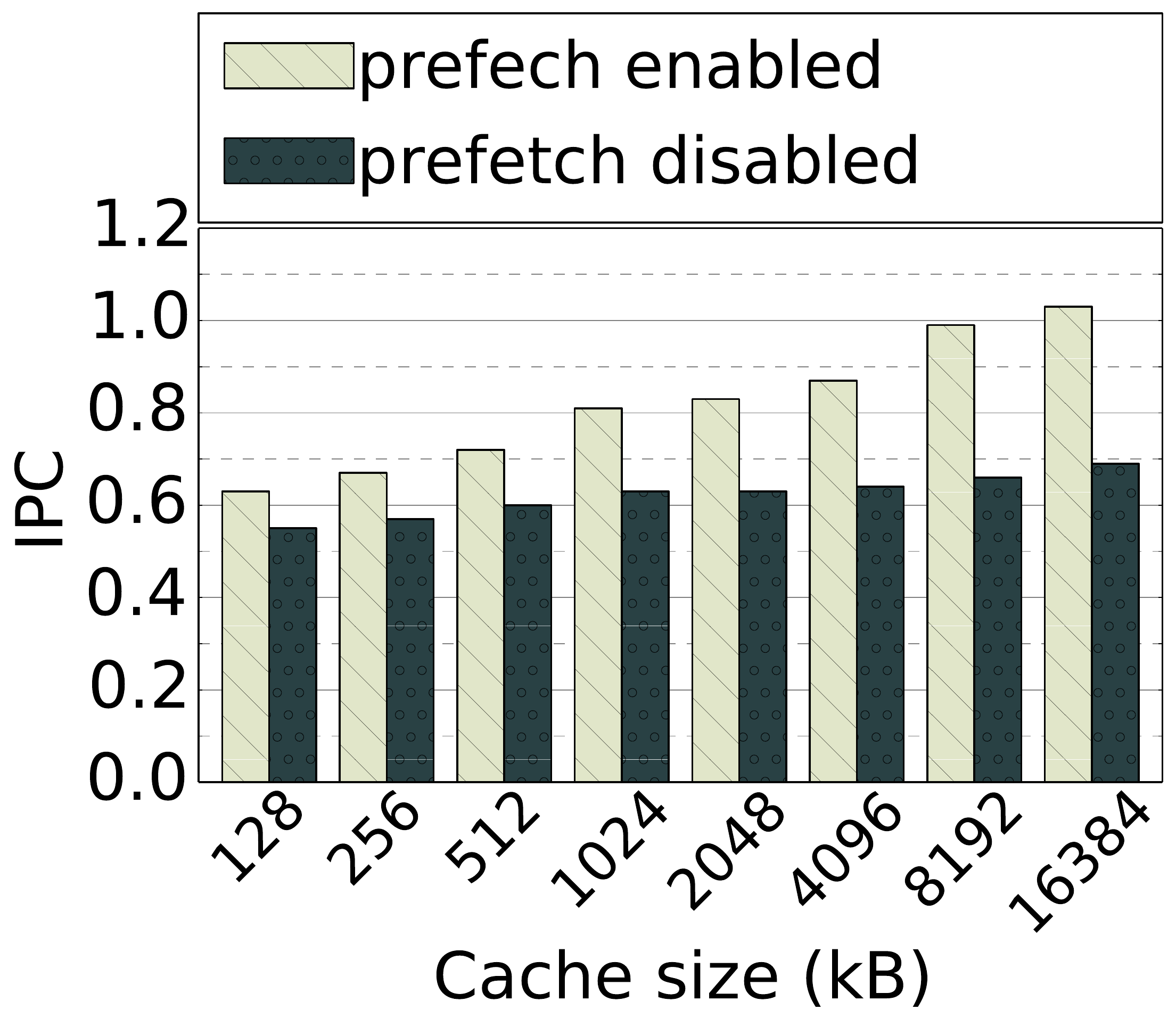}
\vspace{-0.5em}
\caption{IPC with and without prefetching for different cache allocation.}
\label{fig:leslie3d_c}
\end{subfigure}
\hfill
\begin{subfigure}[t]{0.22\textwidth}
\centering
\includegraphics[width=1\textwidth]{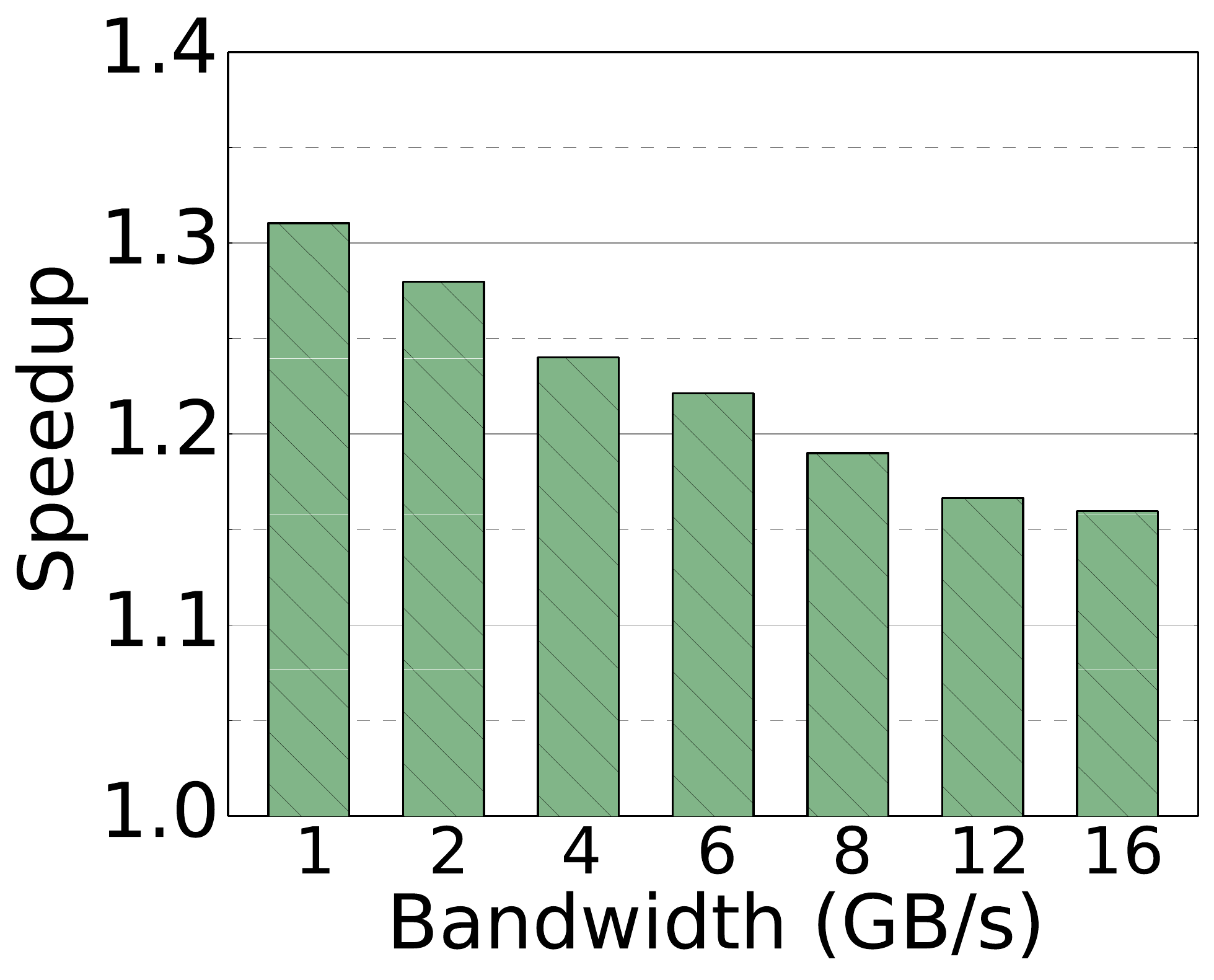}
\vspace{-0.5em}
\caption{Performance improvement from increasing cache allocation to 2MB depending on bandwidth allocation.}
\label{fig:leslie3d_d}
\end{subfigure}
\vspace{-1.5em}
\caption{leslie3d - example of interactions and its impact on performance within a single application.
}
\label{fig:leslie3d_3}
\end{figure*}

In the interest of space we use \textit{leslie3d} as a representative example to illustrate the other pairwise resource interactions and trade-offs, since it is sensitive to all three techniques. Note that the baseline setting will be used for the resources unless specified otherwise. The bandwidth - prefetch interaction manifests in two different ways. Firstly, prefetching typically increases the number of memory accesses and this in turn increases the bandwidth pressure. In the case of \textit{leslie3d} prefetching results in a 15\% increase in the number of memory requests in comparison to the baseline. Note that prefetch misses, i.e. prefetched blocks that are evicted before use, can result in a further increase in pressure on the memory bandwidth. Secondly, the performance improvement from prefetching can be influenced by the bandwidth allocation. The results in Figure \ref{fig:leslie_a} show the performance for different bandwidth allocation with and without prefetching. We make the following observation: 
\vspace{0.2em}

\noindent \textbf{OBSERVATION 3.} \textit{A larger bandwidth allocation can compensate for increased bandwidth demands, due to inaccurate prefetches, leading to increased performance with prefetching.} 
\\
\vspace{-0.5em}

The cache - prefetch interaction also manifests in two main ways. Firstly, the performance loss from reduced cache allocation can be offset if prefetching is effective. Figure \ref{fig:leslie3d_c} shows the IPC for different cache allocations with and without prefetching. The results show that the performance of 128kB allocation with prefetching is better than 512kB allocation without prefetching. 
Secondly, larger cache sizes (if it is used efficiently) can lead to higher speedup from prefetching. Figure \ref{fig:leslie3d_b} shows the performance improvement from prefetching with different cache allocation normalised to the respective cache allocation without prefetching. The results show that prefetching is effective with lower cache allocation and that its effectiveness can increase with additional allocation. The reason for this behaviour is that a larger cache reduces the number of memory accesses which has the same effect as increasing the available bandwidth, i.e. a lower queuing delay, which is more forgiving when there is an increase in memory accesses caused by inaccurate prefetches (in leslie3d there is a 15\% increase in dram accesses caused by prefetching). These results lead to the following observation:
\vspace{0.2em}

\noindent \textbf{OBSERVATION 4.} \textit{ A trade-off can be made between either increasing cache size or enabling prefetching, leading to the same performance, for applications which are performance sensitive to both cache and prefetching.}
\\
\vspace{-0.5em}


As for the cache - bandwidth interaction, a lower bandwidth allocation can result in a larger sensitivity to cache size. Figure \ref{fig:leslie3d_d} shows the performance improvement from increasing the cache allocation from 512kB to 2MB with different bandwidth allocation settings. The results show that performance improvement from additional cache allocation is much higher in low bandwidth allocation settings (see the result for 1GB/s bandwidth allocation). This is because the average cost of a miss is much higher in the case of lower bandwidth allocation. The results also show that a large cache allocation can reduce the performance sensitivity to bandwidth allocation (see the result for 16GB bandwidth allocation). These results lead to the following observation:
\vspace{0.2em}

\noindent\textbf{OBSERVATION 5.}\textit{ A trade-off can be made between either increased cache space or increased bandwidth allocation, for applications which are performance sensitive to both cache and bandwidth.}
\\
\vspace{-0.5em}
   

We now describe how the observed intra-application interactions and trade-offs can be leveraged in the inter-application setting for  multi-programmed workloads. Let us revisit the example of running a simple workload comprising two application (\textit{lbm} and \textit{xalancbmk}) on a dual-core system with 2MB LLC capacity and 16GB/s bandwidth, discussed in Figure \ref{fig:mot_lbm_xalancbmk}.
For achieving the best aggregate performance we expect xalancbmk to get the majority of the cache (nearly 1.75MB), while lbm is given a smaller cache allocation of 256KB. For bandwidth, we would expect lbm to have a large allocation (12GB/s) of the available bandwidth and xalancbmk to get a smaller allocation (4GB/s). This is reflected in Observation 5 about the trade-off between cache and bandwidth where we would prioritize the application that shows the highest sensitivity for the resource. Furthermore, as reflected in Observation 2, we expect prefetching to be more effective for lbm since it has a large allocation of bandwidth. In the case of xalancbmk, prefetching leads to lower performance regardless of the allocation of cache and bandwidth.

\subsection{Potential for coordinated management} 
\label{sec:potential}
\begin{figure}[h!]
\centering
\begin{subfigure}[h]{0.26\textwidth}
\centering
\includegraphics[width=1\textwidth]{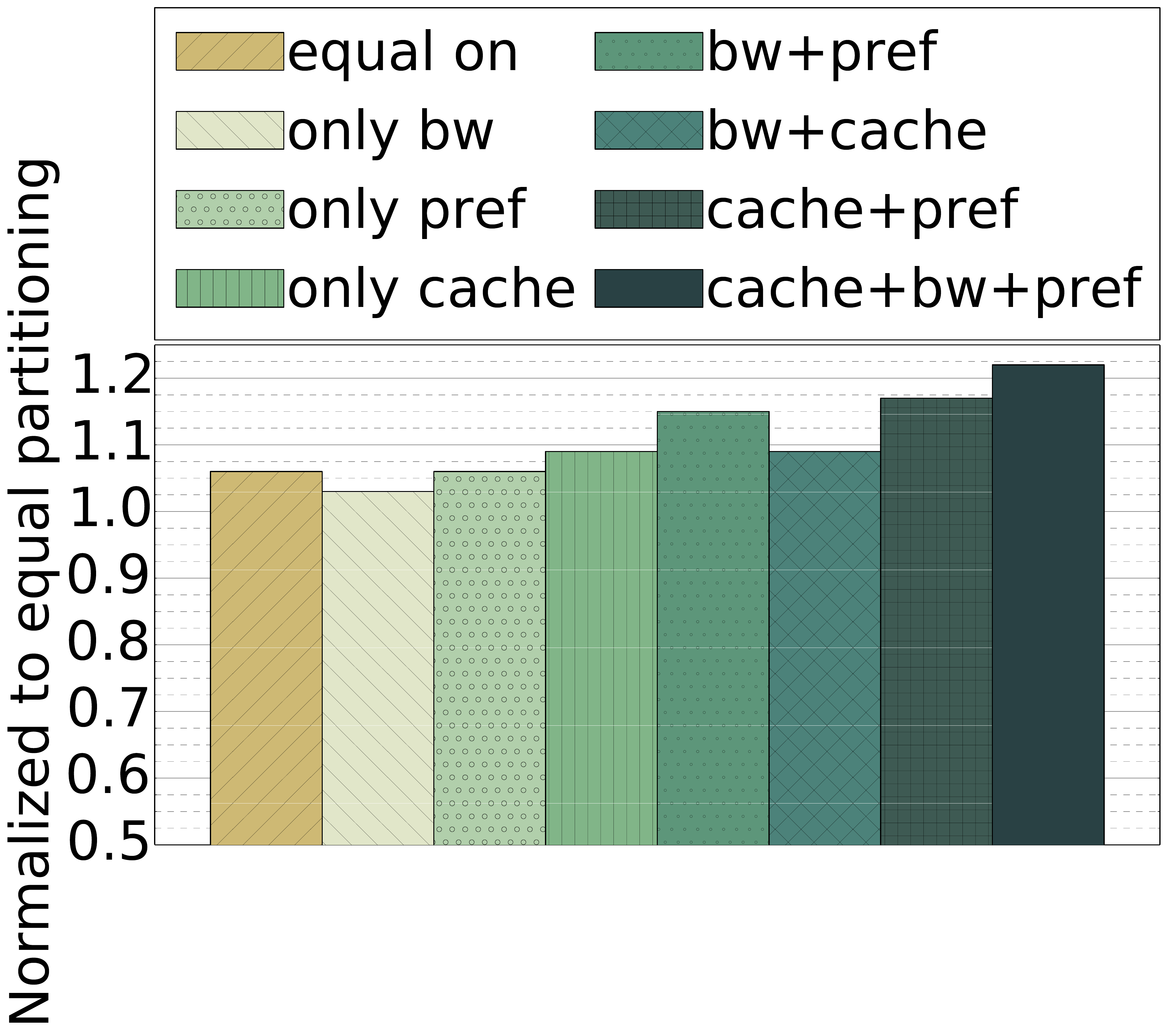}
\caption{Performance potential of using different resource managers. }
\label{fig:static_motivation}
\end{subfigure}
\hfill
\begin{subfigure}[h]{0.2\textwidth}
\centering
\includegraphics[width=1\textwidth]{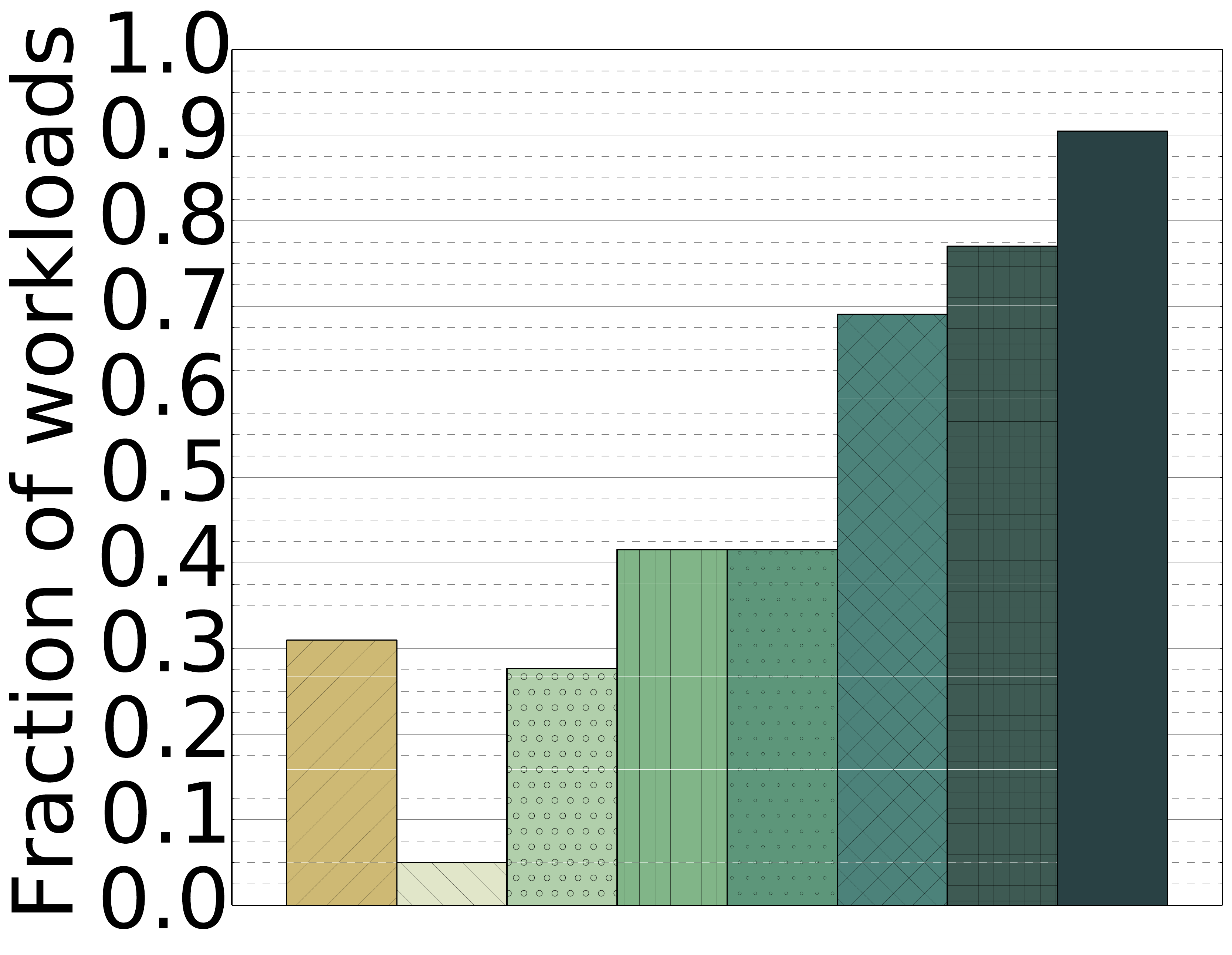}
 \vspace{-1em}
\caption{Fraction, of the workloads, with at least 10\% performance improvement.}
\label{fig:static_numberOfws}
\end{subfigure}
 \vspace{-0.5em}
\caption{Potential for coordinated management measured using 640 random workloads of 4 SPEC CPU2006 applications. Performance is obtained using
an exhaustive search algorithm that evaluates bandwidth settings (2GB/s, 4GB/s, 6GB/s), cache settings (256kB, 512kB, 1024kB) and prefetching settings (active/inactive), in conjunction
, to determine the resource allocation for each application in the workload that maximizes aggregate performance.}
\end{figure}

In order to show the potential for coordinated management of cache, bandwidth and prefetch we run 640 randomly generated workloads each comprising 4 SPEC 2006 CPU applications. We compare the performance of jointly managing all the three resources to other resource managers that only manage a subset of these resources. For this experiment the baseline allocation of cache and bandwidth for each application is 512kB and 4GB/s. We use an exhaustive search algorithm to find the best static configuration (over 1B instructions) for the different resources when running each workload. Figure \ref{fig:static_motivation} shows average (geometric mean) performance with different resource managers normalized to the baseline settings without prefetching. \textit{equal on}, depicts the performance when prefetching is enabled for all applications and improves performance by 6\% while \textit{only pref}, depicts the performance when prefetching is selectively activated and improves performance by 9\%. \textit{cache+bw+pref} results show that coordinately managing cache partitioning, bandwidth partitioning and prefetch throttling improves performance by 5\% compared to the best combination of two techniques (22\% compared to 17\%). 

Figure \ref{fig:static_numberOfws} shows the number of workloads (among the 640 workloads considered) that experience a performance gain of at least 10\% using the different resource managers discussed previously. The results show that
 90\% (597) of the workloads are sensitive to the resource manager that jointly manages all three techniques. A smaller fraction of the workloads are sensitive to resource managers that manage a subset of these techniques (77\% are sensitive to  \textit{cache+pref} resource manager and 69\% are sensitive to the \textit{cache+bw} resource manager). 


In summary, the results from the characterization study demonstrate that around 90\% of applications in the SPEC 2006 CPU suite are sensitive to different resources and that coordinately managing them opens up new possibilities for improving performance by trading resource allocations. Furthermore, jointly managing these resources has the potential to cover a broader range of workloads and outperform resource managers that manage a subset of resources. 
In the next section we will discuss how the proposed resource manager, CBP, determines cache, bandwidth and prefetch settings for different applications in a workload.



\section{CBP resource manager}
\label{sec:descriptionofproposal}

Section \ref{sec:overview} provides an overview of the CBP resource manager. Section  \ref{sec:local_controllers} discusses the individual resource controllers while
%
Section \ref{sec:puttingitalltogether} describes how the coordination mechanism ties the local resource controllers together. Finally, the implementation and overhead of the proposed mechanism is discussed in Section \ref{sec:overhead}.

\subsection{Overview}
\label{sec:overview}

\begin{figure}[b]
\centering
\includegraphics[scale=0.45]{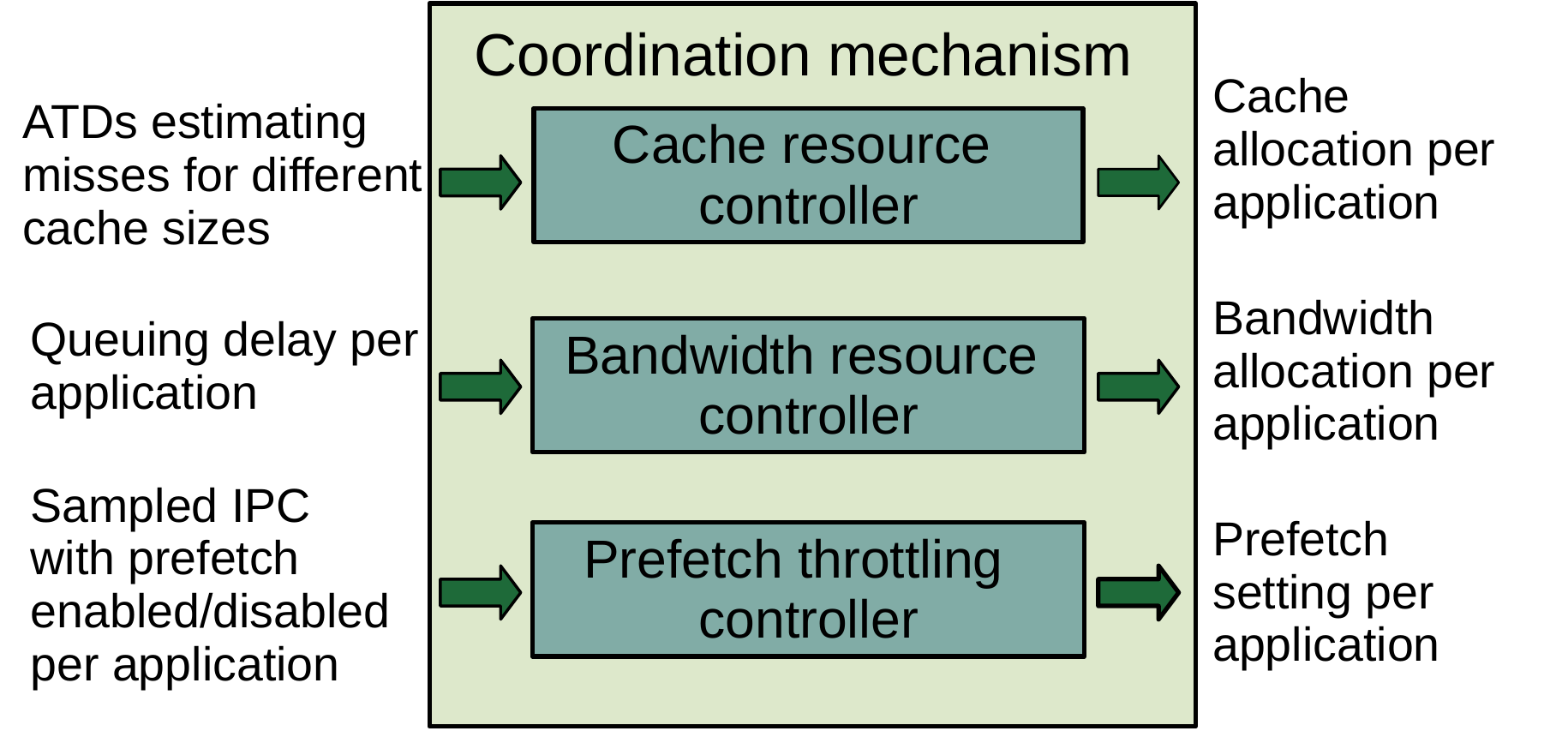}
\vspace{-1em}
\caption{Overview of CBP resource manager.}
\label{fig:cbp_overview}
\end{figure}

CBP is a coordinated mechanism for dynamically managing cache partitioning, bandwidth partitioning and prefetch throttling. The design consists of one local controller for each of the three techniques, and a coordination mechanism, as shown in Figure \ref{fig:cbp_overview}.

The cache allocation controller estimates the number of misses for different cache sizes using auxiliary tag directories (ATDs)~\cite{Qureshi2006} and uses this as input for determining cache allocation. The cache allocation per application is determined such that it reduces the aggregate number of cache misses for the entire workload. The bandwidth allocation controller uses memory request queuing delay experienced by the applications as input and allocates the available bandwidth in proportion to the delay. The bandwidth allocation controller, assigns a larger allocation of the available bandwidth to applications that experience longer queuing delay, and a comparatively lower allocation to those that experience shorter queuing delays. Lastly, the prefetch controller samples IPC with and without prefetching to determine whether the prefetcher should be enabled/disabled for each application. 

The three techniques are dynamically tuned using the coordination mechanism in an iterative manner such that the local controller takes into consideration the decisions taken by the other controllers. 



\subsection{Local resource allocation controllers} 
\label{sec:local_controllers}
A partitioning solution for shared resources like cache and bandwidth typically comprises two components: allocation policy, that determines how a resource is divided among multiple co-running applications, and an enforcement mechanism, that enforces the partitioning decision. Similarly, prefetch throttling involves a policy to determine the best prefetch setting to use and a mechanism implemented in hardware to enforce the setting. In the context of CBP, the policy component is of particular interest because it enables inter-resource trade-offs. We discuss the allocation policy in this section and defer the details of the enforcement mechanism to Section \ref{sec:overhead}.   
\vspace{-0.5em}
\subsubsection{Cache partitioning}
\label{sec:cache}
The cache allocation controller uses the Lookahead algorithm~\cite{Qureshi2006}, to determine cache allocation. In a nutshell, the algorithm computes the utility for each application where utility is the measure of how many additional misses can be reduced with allocation of cache ways. It then computes the number of ways that maximizes the utility for each application (while ensuring this is less than the total number of ways available for allocation). Finally, it compares the utility values for the different applications, determines the application that has the highest utility and assigns the pre-computed number of ways that maximizes the utility for that application. The process repeats, with recomputation of utility for each application and reassignment of available cache ways to the application that has the largest utility, until the rest of the available capacity is distributed. The allocation controller relies on sampled ATDs to estimate, based on past behaviour, the number of misses that can be avoided with additional allocation of cache ways for each application. In order to adapt to an inclusive cache hierarchy, we assign a minimum allocation of cache space 
(\textit{min\_ways}) to all the applications before distributing the remaining capacity. 

\vspace{-0.5em}
\subsubsection{Bandwidth partitioning}
\label{sec:bw}
We propose a bandwidth allocation algorithm, that partitions bandwidth proportional to the memory queuing delay experienced by each application. The pseudo-code for the proposed bandwidth allocation controller is outlined in Algorithm \ref{alg:bw_allocation}. The controller assigns a minimum bandwidth allocation (\textit{min\_bandwidth\_allocation}) for each application in order to avoid unfairly giving a very low allocation to applications with a small queuing delay. The remaining bandwidth is set for distribution among the applications (see line 2). 
The algorithm computes the total queuing delay by summing up the individual queuing delays experienced by each application (line 4) while assigning the minimum allocation to each application (line 5). As the next step, the remaining bandwidth is allocated proportionally to the queuing delay experienced by the application (line 7-9). The queuing delay for each application is obtained by measuring the memory access time for requests from each application. 

\setlength{\textfloatsep}{0pt}
\begin{algorithm}[t!]
\small
 \SetKwInOut{KwIn}{Input}
\SetKwInOut{KwOut}{Output}
\KwIn{A list $queuingDelayPerApplication$}
\KwOut{A list $bandwidthAllocationPerApplication$}
 \emph{\textbf{At time period ${reconfiguration\_interval}$}}\;
  $remainingBandwidth$ = ($totalBandwidth$-$min\_bandwidth\_allocation$*$totalNumberOfCores$
  $totalDelay$ = 0\;
  \For{$i \leftarrow 0$ \KwTo $totalNumberOfCores$$-1$}{
   $totalDelay$ += $queuingDelayPerApplication$[i]\;
   $bandwidthAllocationPerApplication$[i] = $min\_bandwidth\_allocation$\;
  }
  \For{$i \leftarrow 0$ \KwTo $totalNumberOfCores$$-1$}{
      $bandwidthAllocationPerApplication$[i] += ($queuingDelayPerApplication$[i]/$totalDelay$) * $remainingBandwidth$\;
   }
 \caption{Bandwidth allocation controller pseudo-code}
 \label{alg:bw_allocation}
\end{algorithm}

\vspace{-0.1em}
\subsubsection{Prefetch throttling}
\label{sec:pref} 
\setlength{\textfloatsep}{0pt}
\begin{algorithm}[b!]
\small
 \SetKwInOut{KwIn}{Input}
\SetKwInOut{KwOut}{Output}
\KwIn{Two lists $ipcWithPrefetchingActive$, $ipcWithPrefetchingInactive$}
\KwOut{A list $prefetchSettingPerCore$}
 \emph{\textbf{At time period ${prefetch\_interval}$}}\;
  \For{$i \leftarrow 0$ \KwTo $totalNumberOfCores$$-1$}{
  \eIf{($ipcWithPrefetchingActive$[i]/$ipcWithPrefetchingInactive$[i]) > $speedup\_threshold$}{
   $prefetchSettingPerCore$[i] = 0\;
  }{
   $prefetchSettingPerCore$[i] = 1\;
  }
  }
  
 \caption{Prefetch throttling controller pseudo-code}
 \label{alg:pref_throttling}
\end{algorithm}
\vspace{-1em}

The prefetch throttling policy determines the best prefetcher settings for each application. The pseudo-code for prefetch throttling controller is outlined in Algorithm \ref{alg:pref_throttling}. The algorithm considers two possible settings -- prefetcher enabled and prefetcher disabled -- but can easily be extended to support other aggressiveness settings as well. The algorithm uses the sampled IPC values, for each application obtained, with different prefetcher setting over a sample period 
(\textit{prefetch\_sampling\_period}) as input. 
The algorithm first computes the speedup from prefetching for each application using the sampled IPC values. If the speedup is below a threshold 
(\textit{speedup\_threshold}) the prefetcher is deactivated for the next prefetch interval (\textit{prefetch\_interval}) (line 3-4). If the speedup is above the threshold prefetching is activated for the next prefetch interval (line 6). The prefetch throttling controller is generic enough to support any type of prefetcher.


\subsection{Coordination Mechanism}
\label{sec:timing}
The goal of the coordination mechanism is to ensure that each local controller takes into account the decisions taken by other controllers. This is necessary to exploit the trade-offs outlined earlier in Section \ref{sec:motivation}. There are two essential tasks carried out in order to establish this: i) controller prioritization and ii) inter-controller interaction.

\textbf{Controller prioritization:} Since the decision taken by one controller has the potential to influence those taken by others it is important to establish priority among the different local controllers since that determines the order in which local controllers make allocation decisions. 
In CBP, the highest priority is given to the cache allocation controller, which first makes the allocation decision. The rationale is that avoiding a memory access is typically more effective, for reducing average memory access time, than lowering the memory access penalty. Next, priority is given to the bandwidth allocation controller since our characterization results show that applications are comparatively more sensitive to bandwidth than to prefetching. 
The least priority, is given to the prefetch throttling controller. This is because it is important that the prefetcher setting is determined based on the current allocation, of cache and bandwidth, since prefetching can have a negative impact on performance if the bandwidth allocation is insufficient.

\textbf{Inter-controller interaction: } 
\begin{figure}[b]
\centering
\includegraphics[scale=0.42]{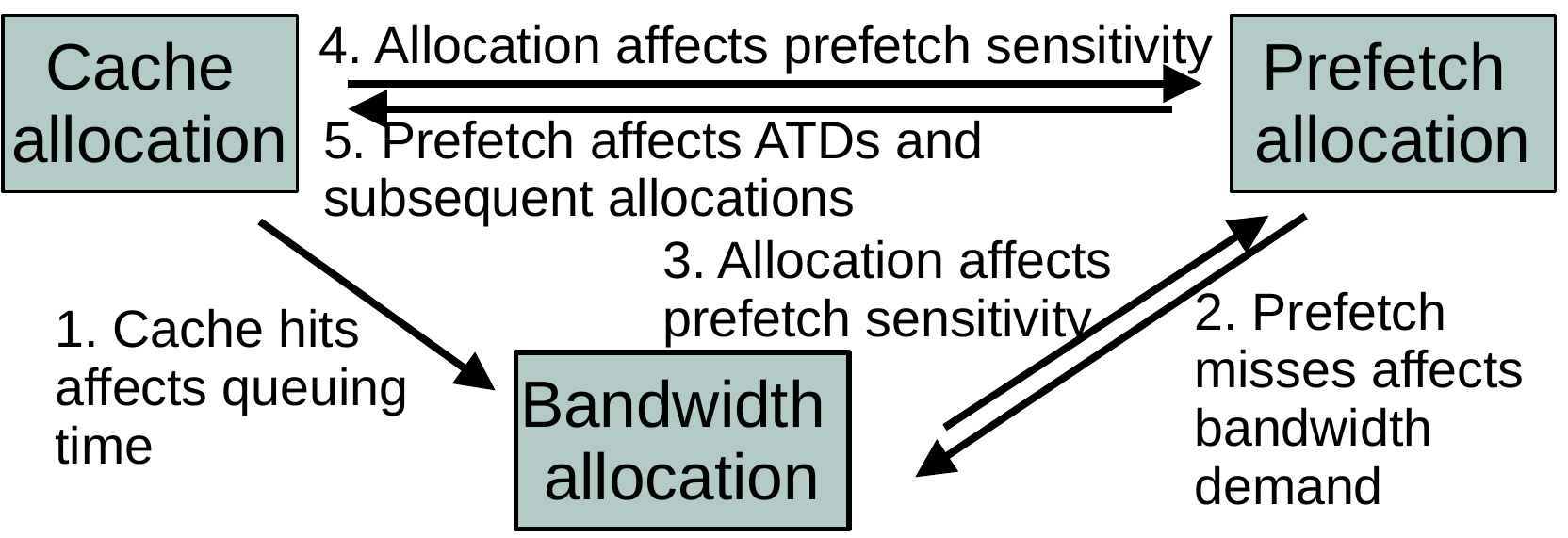}
\vspace{-1.5em}
\caption{Interactions among the different resource allocation controllers in CBP .}
\label{fig:cbp_interactions}
\end{figure}
Figure \ref{fig:cbp_interactions} provides an overview of the interactions that happens between the different resource allocation controllers. Firstly, we describe how the bandwidth allocation controller decisions takes into account the decisions made by the cache and the prefetch controller.
The bandwidth allocation controller, makes decisions based on the queuing delay of each application which is affected by the number of memory accesses. The cache allocation controller through a larger cache allocation can reduce the number of memory accesses (Interaction \#1). This leads to a lower bandwidth allocation for applications that can efficiently use the cache.
The prefetch throttling controller, influences bandwidth allocation decision mainly through prefetch misses (prefetched data that is not used) (Interaction \#2). This is because prefetch misses lead to more memory requests and potentially a higher queuing delay. 

Next, we describe how the prefetch throttling controller takes into account the decisions made by the other controllers.
The prefetch throttling controller, makes decisions by sampling the IPC with different prefetcher setting over a specific interval. The sampled IPC values, used to determine the prefetcher setting, reflects the effect of cache and bandwidth allocation decisions made by the respective resource controllers (Interaction \#3-4). Finally, we describe how the cache allocation controller is affected by the prefetch throttling controller (Interaction \#5). If an application benefits from prefetching this reflects on the hit and miss count values monitored in the ATDs. Since the cache allocation, is computed based on the counter values observed in the ATD, this end up affecting the subsequent cache allocation decision, resulting in a smaller cache allocation for prefetch sensitive applications.

\begin{figure}[h]
\centering
\includegraphics[scale=0.47]{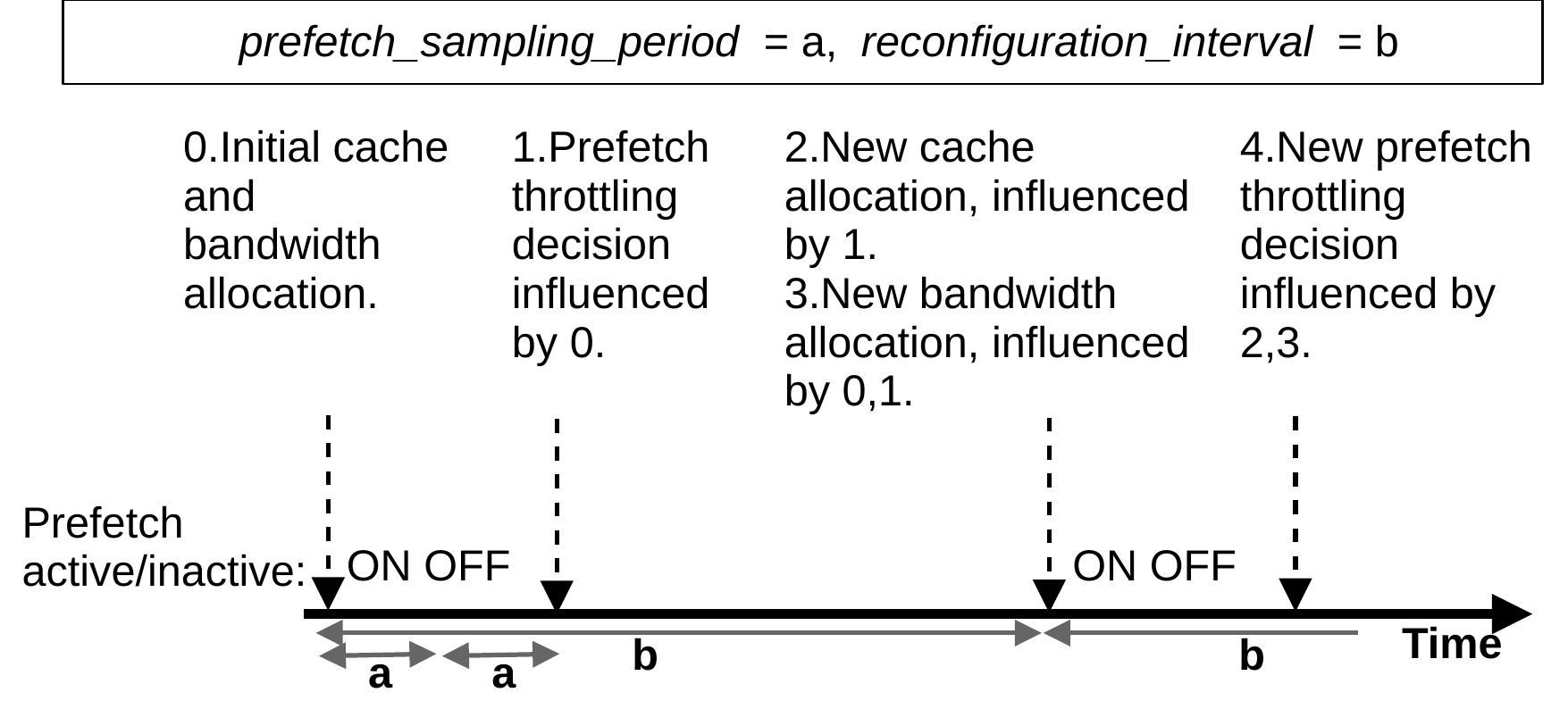}
\vspace{-1em}
\caption{Timing and interactions of CBP resource manager.}
\label{fig:cbp_timing}
\end{figure}

\textbf{Putting it all together: }We discuss the timeline showing when the different resource allocation controllers are invoked and how they interact with each other, in an iterative manner, using an example illustrated in Figure \ref{fig:cbp_timing}.  The three resource controllers, are invoked after a specific interval that we refer to as the \textit{reconfiguration\_interval} in the sequence shown in the figure. First cache and bandwidth are equally partitioned among all applications at time 0, since information about misses and queuing delay is initially unavailable, as shown in Step 0. This is followed by sampling the IPC of applications with different prefetch settings for a specific interval (twice the \textit{prefetch\_sampling\_period}) as shown in Step 1. 
Based on the sampled IPCs the prefetch throttling controller determines the appropriate prefetcher setting for each prefetcher for the current \textit{reconfiguration\_interval}. Cache and bandwidth allocation controllers are again invoked after the \textit{reconfiguration\_interval}, as shown in Step 2. A cache allocation decision, for the next interval is influenced by the number of hits and misses observed in the previous interval. The ATD values will be halved after each reconfiguration, in order to be sensitive to changes in the last time interval while the per application queuing delays are accumulated with those from the previous interval. The bandwidth allocation decision shown in Step3, is influenced both by the cache allocation and prefetcher setting in the previous interval as discussed previously. Finally, the prefetcher throttling controller shown in Step 4 is influenced by the new cache and bandwidth allocation. The interactions among the different resource allocation controllers, take place over multiple iterations, and is the key to finding an effective solution. 

\label{sec:puttingitalltogether}
 
\vspace{-1em}
\subsection{Implementation}
\label{sec:overhead}

The computational overhead of CBP resource management is low since the design uses heuristics to guide the allocation decisions instead of exhaustively evaluating the different possible allocations. 

The cache allocation controller needs hardware support in order to estimate the number of misses with different cache sizes. We use sampled ATDs~\cite{Qureshi2006} as discussed previously to compute the effect of different cache allocations on the misses.
When enforcing cache partitioning there is an overhead associated with invalidations due to reconfiguration decisions. This is modelled faithfully by invalidating the addresses and re-fetching them when accessed, this includes the latency and impact on bandwidth from accessing memory. We have used the enforcement mechanism proposed by Holtryd et al.~\cite{holtryd2020} since it is suitable for a modern tile-based CMP and is both fine-grained and locality aware. The enforcement mechanism uses per-core Cache Bank Tables (CBTs), where mappings between addresses and banks are recorded. When a request needs to access the LLC, the CBT is used to identify the cache bank that the address is mapped to. 
Inside each bank, way partitioning hardware divides the capacity. 
The enforcement results in a partition granularity of 32kB on our system, see Section \ref{sec:methodology}. We incur hardware cost for implementing ATDs and cache partition enforcement ~\cite{Qureshi2006,holtryd2020}. 

The bandwidth partition enforcement is done in likeness with Intel Memory Bandwidth Allocation (MBA) technology~\cite{inteldevelopermanual_2021,mba_article} which is commercially available. The solution uses delays as a way to allocate the bandwidth. An application with a high delay has a low allocation, and experiences a longer queuing delay for each memory access. 
In our solution the additional delay is added after the LLC, instead of after the L2, as in the original proposal.  

The overhead of prefetch-throttling comes from sampling an application with different prefetcher settings. This is because deactivating prefetching for an application can be detrimental for its performance, especially when prefetching is effective. Likewise, it is detrimental to turn it on prefetching (for a sample period) for an application whose performance is hurt by prefetching.

\vspace{-0.5em}
\section{Experimental Methodology}
\label{sec:methodology}
\label{sec:experimentalsetup}
\subsection{Simulated Architecture}
\label{sec:platform}
We evaluate our proposal on a 16-core tiled CMP architecture modeled using the Sniper Simulator \cite{Carlson:2014:EHM:2658949.2629677}. Each tile has an out-of-order (OOO) core with a private L1 data and instruction cache, a unified private L2 cache and an LLC bank of 512KB. The cache latencies assumed have been modelled using CACTI 6.5 \cite{4408241}. Details about the baseline architecture are shown in Table~\ref{table:config}. A sensitivity study is provided for the CBP parameters in Section~\ref{sec:sensitivitystudy}.

\begin{scriptsize}
\begin{table}[h!]
  \centering
  \small
  \scalebox{0.89}{
  \begin{tabular}{|l|l|}
    \hline
    \textbf{Cores} & 16 cores, x86-64 ISA, 4GHz, OOO, \\   &
    Nehalem-like, 128 ROB entries, dispatch width 4 \\
    \hline
    \textbf{L1 caches} & 32KB, 8-way set-associative, split D/I, \\   &1-cycle latency \\
    \textbf{L2 caches} & 128KB private per-core, 8-way set-associative,\\   & inclusive, 6-cycle data and 2-cycle tag latency \\
    \textbf{LLC} & 512KB per-tile, 16-way set-associative, inclusive,\\   & 9-cycle data and 2-cycle tag latency, LRU \\
    \hline
    \textbf{Coherence protocol} & MESIF-protocol, 64 B lines, in-cache directory \\
    \hline
    \textbf{Global NoC} & 4x4 mesh, 4-cycles hop latency \\   &
    (3-cycle pipelined routers, 1-cycle links) \\
    \hline
    \textbf{Memory controllers} & 4 MCUs, 1 channel/MCU, latency 80 ns,\\   &16GB/s per channel \\
    \hline
    \textbf{Prefetcher} & stride-based, located in L2, 4 prefetches \\&  stop at page boundary, 8 flows/core \\
    \hline
    \hline
    \textbf{CBP parameters} &  
    \textit{reconfiguration\_interval}=10ms \\ &
    \textit{prefetch\_sampling\_period}=0.5ms, \\ 
    &\textit{speedup\_threshold} = 1.05, \\   
    &\textit{prefetch\_interval}=10ms, \\
    &\textit{min\_bandwidth\_allocation}=1, \textit{min\_ways}=4 \\ 
    \hline
  \end{tabular}}
  \caption{Configuration of the simulated 16-core tiled CMP. }
   \vspace{-1em}
  \label{table:config}
\end{table}
\end{scriptsize}

\vspace{-0.5em}
\subsection{Methodology}
\label{sec:classification}
We use the entire SPEC CPU2006 suite in our evaluation. The applications are in the format of whole program pinballs \cite{Sherwood:2002:ACL:605397.605403}. 
We create 14 workload mixes (each comprising 16 applications) by randomly selecting applications from the entire SPEC CPU2006 suite.
Details about workload mixes are presented in Table~\ref{table:mixes_100}.

\begin{scriptsize}
\begin{table}[t!]
  \centering
  \small
  \scalebox{0.82}{
  \begin{tabular}{|l|l|l|}
    \hline
    \textbf{w\#} & \textbf{Types} & \textbf{Benchmarks} \\
    \hline
    w1 & 4CS-BS-PS,5CS-BS,3BS-PS,3CS,1BS & xalancbmk(xa),gromacs(gr),libquantum(li)(2)\\  
    && h264ref(h2),zeusmp(ze),tonto(to),soplex(so),\\ && lbm(lb),perlbench(pe),calculix(ca),milc(mi)\\ && sphinx3(sp),bwaves(bw),gobmk(go),gamess(ga) \\ 
    \hline
    w2 & 3CS-BS-PS,5CS-BS, & lb,to,pe,go,gcc(gc),mi,li(2),namd(na), \\ & 5BS-PS,2CS,1BS  & 
    h2,cactusADM(cac),ze(2),ca,so,astar(as) \\
    \hline
    w3 & 6BS-PS,CS,5BS,4I & 
    bw(2),povray(po)(2),sjeng(2),sp(2),na(2),ze, \\ && GemsFDTD(Ge),cac,li,mi,wrf(wr)\\
        \hline
    w4 & CS-BS-PS,2CS-BS,5BS-PS,3CS,2BS,3I & po,bw(2),h2,sjeng(sj),li(2),gr,na,mi(2),as,Ge, \\ && ga,wr,lb \\
        \hline
    w5 & 5CS-BS-PS,10CS-BS,BS-PS & dealII(de),omnetpp(om)(2),go(2),hmmer(hm),xa\\ &  & 
    leslie3d(le),bzip2(bz)(2),gc,so,mcf(mc),pe,ca(2)\\
        \hline
    w6 &  3CS-BS-PS,5CS-BS,4BS-PS,2CS,2BS & sp,bw(2),h2,om,li,gr,go,mi(2),as,hm,ga,le,lb,ca \\ 
        \hline
    w7 & 2CS-BS-PS,2CS-BS,3BS-PS,5CS,4I & po(2),to,sj,h2(2),na,lb(2),ze(2),gr,Ge,as,wr,ga \\ 
        \hline
    w8 & 4CS-BS-PS,4CS-BS,2CS-PS,3BS-PS & de,bw(3),xa,mi(3),om,li(2),bz,go,so,hm,pe \\ & 3BS \\ 
        \hline
    w9 & 2CS-BS-PS,5CS-BS,2BS-PS,3CS,BS,2I & gc,po,to,hm,sj,h2,bz,ze,gr,so,Ge,as,pe,wr,ga,cac \\ 
        \hline
    w10 & 2CS-BS-PS,3CS-BS,6BS-PS,CS,2BS,2I & sj,bw(2),de,na,li(2),om,ze,mi(2),xa,Ge,bz,wr,gc\\ 
        \hline
    w11 & 2CS-BS-PS,4CS-BS,4BS-PS,CS,2BS,3I & po,om,sj,go,na(2),le,ze,xa,Ge,bz,wr,ca,sj,sp,gc \\ 
           \hline
    w12 & 6CS-BS-PS,8CS-BS,2CS & de,to,go,h2(2),hm,gr,xa,as(2),bz,ga,gc,lb,so,ca\\
           \hline
    w13 & 3CS-BS-PS,2CS-BS,4BS-PS,4CS,3I & to,po,h2,sj,gr,na,as,ze,ga,Ge,lb(2),li,to,mi,wr \\
           \hline
    w14 & 5CS-BS-PS,2CS-BS,5BS-PS,CS,BS,2I & de,bw,go,po,hm,na,xa,ze,so,Ge,mc,li, \\&& pe,mi,ca,wr \\
      \hline
  \end{tabular}}
  \caption{16-core workload. }
  \label{table:mixes_100}
\end{table}
\end{scriptsize}

We fast-forward for 16B instructions (in total) and then carry out detailed simulation until all benchmarks have completed at least 500M instructions. Statistics are reported based on the detailed simulation of 500M instructions. After this period the applications continue to run and compete for resources to avoid having a lighter load on long running applications. The methodology is in line with earlier works \cite{Sanchez2011,Beckmann2013,Manikantan2012}.

\vspace{-0.5em}
\subsection{Metrics} 
\label{sec:performancemetrics}
We report normalized weighted speedup over baseline for each workload. 
This is computed by $\frac{1}{N}\sum_{i=1}^{N}\frac{IPC_{i,RM}}{IPC_{i,baseline}}$
, in order to evaluate system performance for multi-programmed workloads. 
RM refers to the system with a resource manager that manages cache, bandwidth and prefetcher settings and the baseline refers to a system with unpartitioned cache and bandwidth and without prefetching.


We also report average normalized turnaround time (ANTT) for each workload since this is a user-oriented performance metric which shows fairness. 
ANTT is given by $\frac{1}{N}\sum_{i=1}^{N}\frac{CPI_{i,RM}}{CPI_{i,baseline}}$ 

\vspace{-0.5em}
\subsection{Comparison}
\label{sec:comp}
Table \ref{table:designs} shows the resource managers we evaluate, in addition to CBP, and the corresponding settings they use for cache, bandwidth and prefetching. The baseline configuration represents a system with unpartitioned cache, unpartitioned bandwidth and with prefetching disabled. \textit{equal off} configuration represents  a system where cache and bandwidth are equally partitioned and prefetching is disabled. 
\textit{only cache} represents the configuration where cache is partitioned as described in Section \ref{sec:cache}, while bandwidth is unpartitioned and with prefetching disabled. Likewise, \textit{only bw} represents the configuration where bandwidth is partitioned as described in Section \ref{sec:bw} while the cache is unpartitioned and with prefetching disabled. As for \textit{only pref}, prefetch throttling is performed as described in Section \ref{sec:pref} while cache and bandwidth remains unpartitioned. The resource managers that jointly manage two out of the three resources (\textit{bw+perf, bw+cache, cache+perf}) in a coordinated manner can leverage a subset of the interactions described in Section \ref{sec:timing}. We also compare against \textit{CPpf}~\cite{Xiao2019}, a recently proposed technique for jointly managing prefetching and cache partitioning. In \textit{CPpf}, prefetch friendly applications are allocated small partition sizes (because the benefit from prefetching can offset the performance drop from small allocation) while the rest of the cache is allocated to the non prefetch friendly applications. In our implementation, we give minimum allocation to the prefetch friendly applications and use UCP (see Section~\ref{sec:cache}), to partition the remaining capacity among the non prefetch friendly applications. We use UCP and per application partitioning, in order to not put \textit{CPpf} at a disadvantage in comparison to other schemes. Finally, CBP jointly manages all the three techniques dynamically.

\begin{scriptsize}
\begin{table}[t!]
  \centering
  \small
  \scalebox{0.89}{
  \begin{tabular}{|l|l|l|l|}
    \hline
    \textbf{RM} & \textbf{cache setting} & \textbf{bandwidth setting} & \textbf{prefetch setting}\\
    \hline
   \hline
   baseline & unpartitioned &
   unpartitioned & disabled \\
   \hline
equal off & equal &
   equal & disabled \\
   \hline
   only cache & dynamic (see 3.2.1) &
   unpartitioned & disabled \\
   \hline
   only bw & unpartitioned &
   dynamic (see 3.2.2) & disabled \\
   \hline
   only pref & unpartitioned &
   unpartitioned & dynamic (see 3.2.3) \\
   \hline
   bw+pref & unpartitioned &
   dynamic & dynamic \\
   \hline
   bw+cache & dynamic &
   dynamic & disabled \\
   \hline
   cache+pref & dynamic &
   unpartitioned & dynamic \\
   \hline
   CPpf & dynamic &
   unpartitioned & enabled \\
   \hline
   CBP & dynamic &
   dynamic & dynamic \\
   \hline
  \end{tabular}}
  \caption{Configurations evaluated }
  \label{table:designs}
\end{table}
\end{scriptsize}

\vspace{-0.5em}
\section{Evaluation}
\label{sec:evaluation}

We first compare CBP to other resource managers that manage a subset of resources. Next, we carry out sensitivity analysis for the different design parameters, to understand its impact on the performance of CBP.

\vspace{-0.5em}
\subsection{CBP Performance Analysis}
\label{sec:16core}
\begin{figure}[b!]
\centering
\centering
\includegraphics[width=0.49\textwidth]{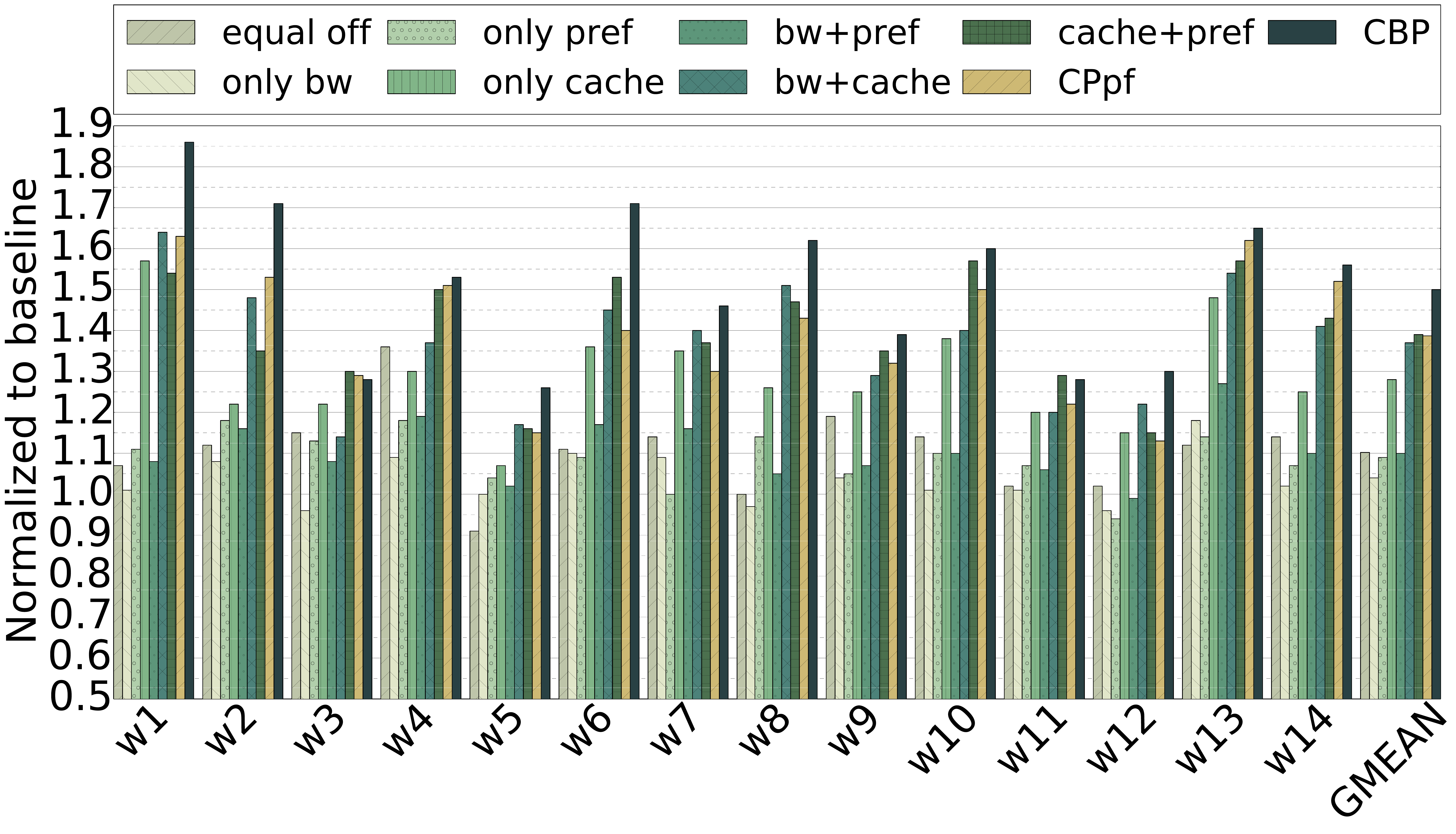}
\vspace{-3em}
\caption{Performance results, shows normalized weighted speedup over baseline.}
\label{fig:results_mixOf16}
\end{figure}

Figure \ref{fig:results_mixOf16} shows normalized weighted speedup for each of the 14 workloads with the bars representing the different resource managers. We use normalized weighted speedup as a measure of performance for the entire workload.\textit{equal off} improves performance in 12 of the mixes and improves performance by 10\% on average over the baseline. \textit{only bw} improves performance in 7 of the mixes and on average by 4\% over the baseline. \textit{only pref} improves performance for 12 workloads and provides an average improvement of 9\%. \textit{only cache} improves performance for all the workloads and provides an average improvement of around 28\%.

Coordinated management of bandwidth partitioning and prefetch throttling (\textit{bw+pref}) leads to higher performance in comparison to the baseline in 12 workloads and an average overall improvement of 10\%. Coordinated management of bandwidth and cache partitioning (\textit{cache+bw}) improves performance across all workloads and provides an average performance improvement of 37\% (up to 64\%). Coordinated cache partitioning and prefetch throttling (\textit{cache+pref}) improves performance across all workloads on average by 39\% (up to 57\%). \textit{CPpf}, cache partitioning influenced by prefetching, improves performance by 39\% (up to 63\%).

Among the resource managers that performs coordinated management of two resources, \textit{cache+pref} and \textit{CPpf} achieve the best performance.
The results also show that the improvement achieved with coordinated management of two techniques, is larger than summing the improvements from individual techniques. This shows that coordinated management helps exploit synergistic interactions among the different techniques, which cannot otherwise be leveraged. Finally, CBP turns out to be the best performing coordinated resource manager in 14 of the 15 workloads and provides an average improvement of 50\% (up to 86\%). 
CBP improves performance by an additional 11\% in comparison to the best performing resource manager that does coordinated management of two techniques, as well as state-of-the-art. In one workload, w3, CBP achieves slightly lower performance (2\%) in comparison to \textit{cache+pref}. This is because bandwidth partitioning is not very effective for this specific workload.

\begin{figure}[t]
\centering
\includegraphics[width=0.49\textwidth]{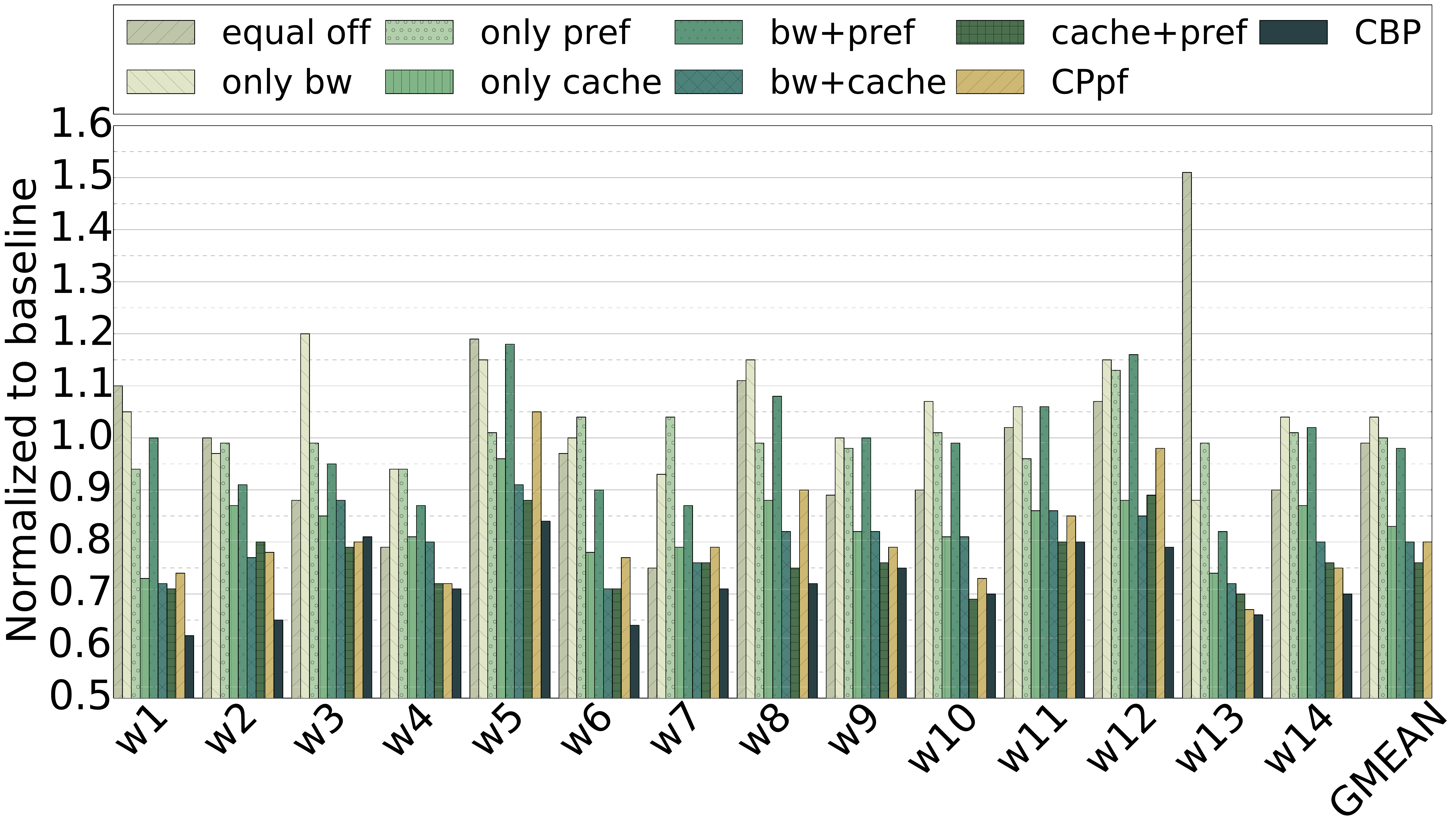}
\vspace{-3em}
\caption{Fairness results, shows average normalized turnaround time (ANTT) over baseline, where lower is better.}
\label{fig:fairness_cbp}
\end{figure}

Figure \ref{fig:fairness_cbp} shows the average normalized turnaround time which shows the fairness of the different resource managers. Note that a lower value signifies greater fairness. On average, CBP shows 27\% better fairness than the baseline and 4\% better fairness than the best combination of two techniques, \textit{cache+pref}. \textit{cache+pref} has 4\% better fairness than \textit{CPpf}. 

\begin{figure}[t!]
\centering
\includegraphics[width=0.48\textwidth]{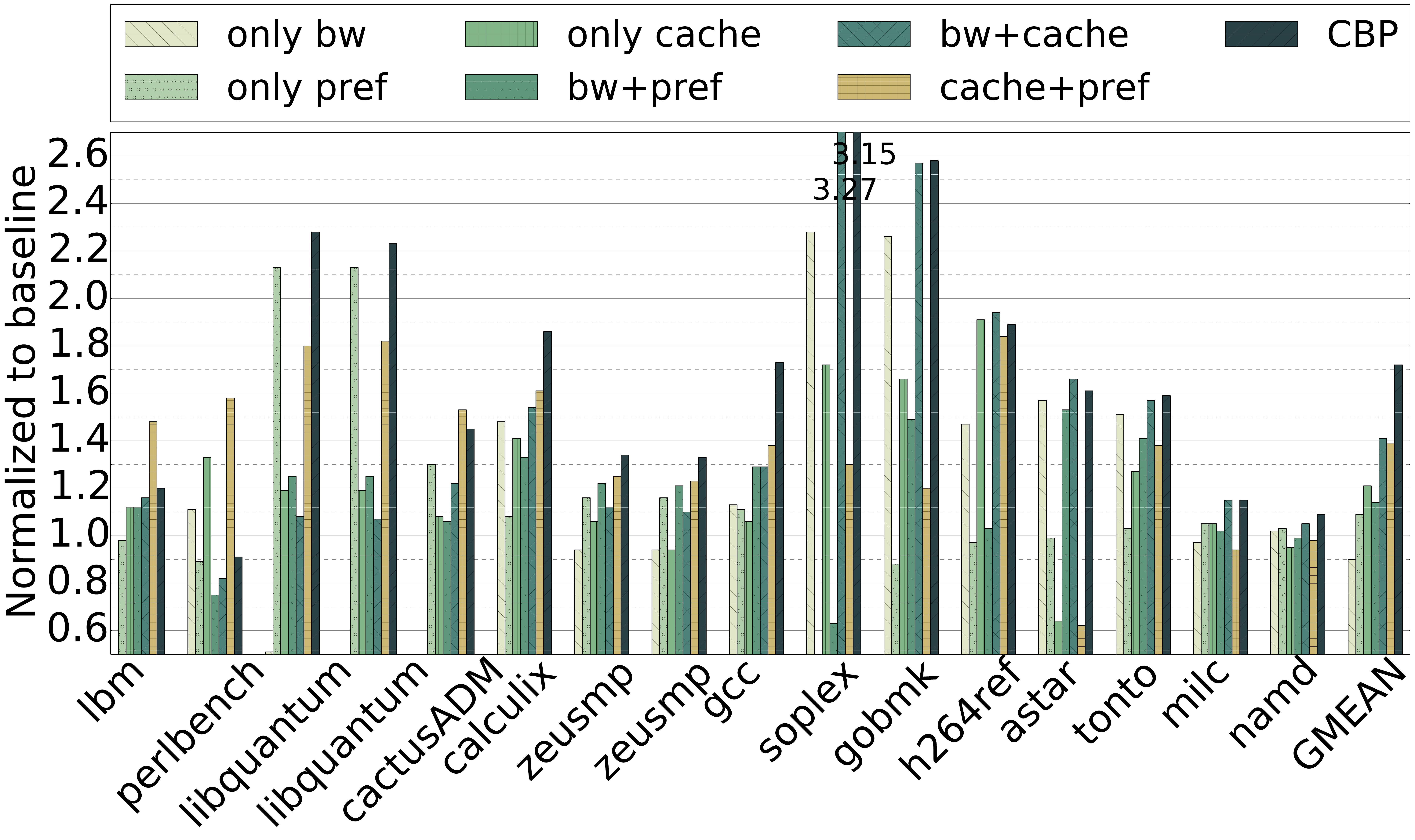}
\vspace{-2.5em}
\caption{Results for workload 2. 
}
\label{fig:results_w2}
\end{figure}

\textbf{Case study:} We investigate the performance of a single workload in detail to understand how CBP improves performance in comparison to resource managers that manage a subset of the techniques. Figure \ref{fig:results_w2} shows the IPC for individual applications in a specific workload (w2) normalized to the baseline IPC. 
We have classified the applications in this workload into two groups. Group 1 comprises applications, from lbm to gcc (in the figure), for which the \textit{cache+pref} resource manager performs better than the \textit{bw+cache} resource manager. Group 2 comprises the rest of the applications in the workload, from soplex to namd, where the \textit{bw+cache} resource manager performs better than the \textit{cache+pref} resource manager. \textit{cache+pref} resource manager provides the best performance for applications in group 1 because the applications are comparatively more memory intensive and get a larger share of the available bandwidth using this resource manager. Applications in group 2 benefit from bandwidth partitioning since they then get a fair bandwidth share, and in addition are not sensitive to prefetching. When we perform coordinated management of all the three resources, we would ideally prefer to have allocation decisions made by \textit{cache+pref} resource manager for applications in group 1 and \textit{bw+cache} resource manager for applications in group 2. With CBP, some applications in group 1 end up with a lower allocation of bandwidth (compared to \textit{cache+pref}) which hurts their performance (see lbm, perlbench, cactusADM) while the rest of the applications in the group see a performance improvement from getting the right amount of the allocation. For the applications in group 2, CBP manages to match the performance of \textit{bw+cache} resource manager.  
In summary, CBP enables better trade-offs resulting in a solution that improves overall performance for the workload and outperforms other resource managers that only manage a subset of the techniques.

\vspace{-1em}
\subsection{Sensitivity analysis}
\label{sec:sensitivitystudy}
We investigate the sensitivity of CBP to different design parameters in this section. 

\vspace{-0.5em}
\subsubsection{Impact of reconfiguration interval}
\label{sec:recon}
\begin{figure}[t!]
\begin{subfigure}[t]{0.12\textwidth}
\centering
\includegraphics[width=0.77\textwidth]{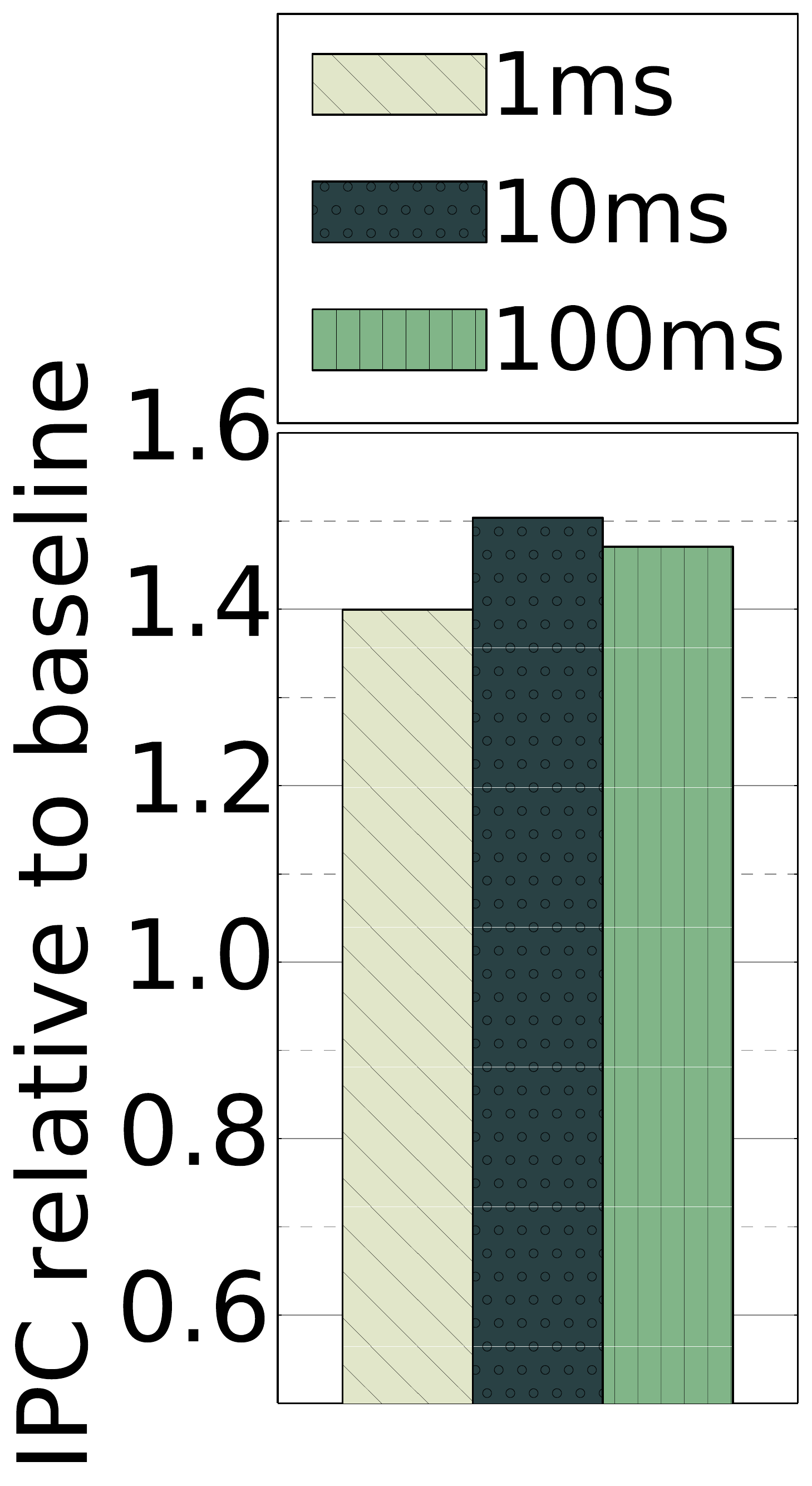}
\caption{Sensitivity for reconfiguration period.}
\label{fig:sen_recon}
\end{subfigure}
\hfill
\begin{subfigure}[t]{0.10\textwidth}
\centering
\includegraphics[width=0.9\textwidth]{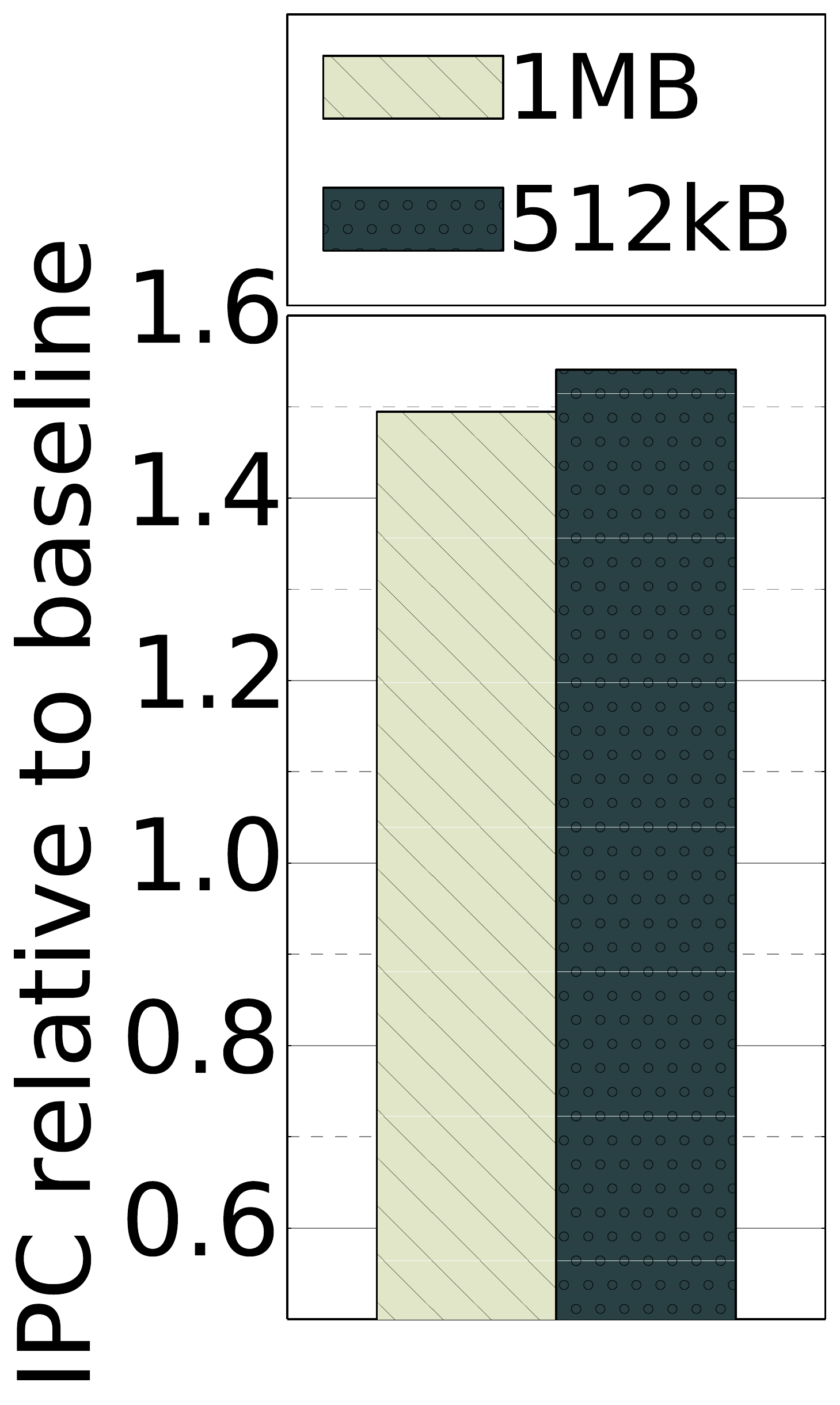}
\caption{Sensitivity for larger cache size.}
\label{fig:sen_cache_size}
\end{subfigure}
\hfill
\begin{subfigure}[t]{0.09\textwidth}
\centering
\includegraphics[width=\textwidth]{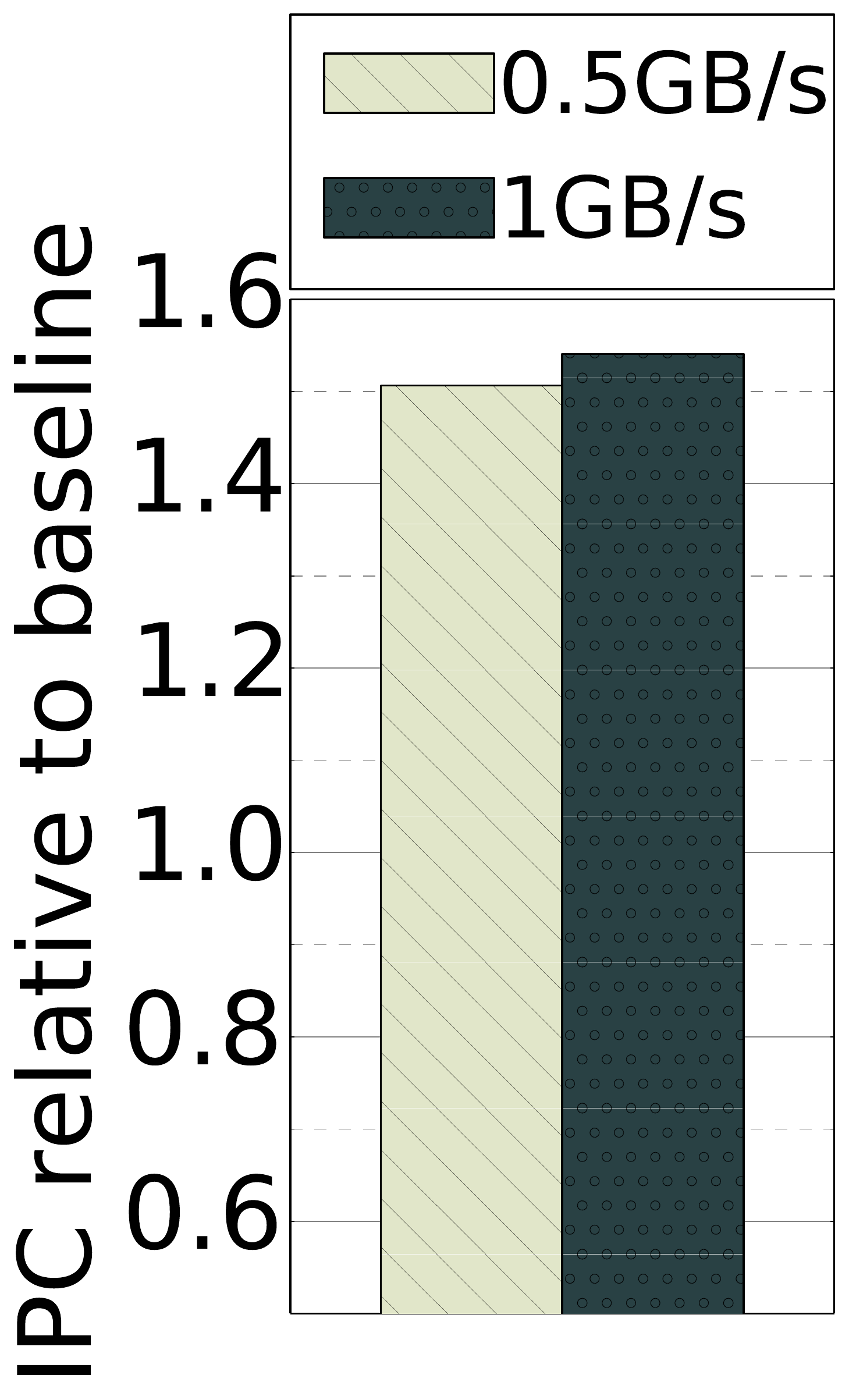}
\caption{Sensitivity for minimum bandwidth allocation.}
\label{fig:sen_min_bw}
\end{subfigure}
\hfill
\begin{subfigure}[t]{0.12\textwidth}
\centering
\includegraphics[width=0.75\textwidth]{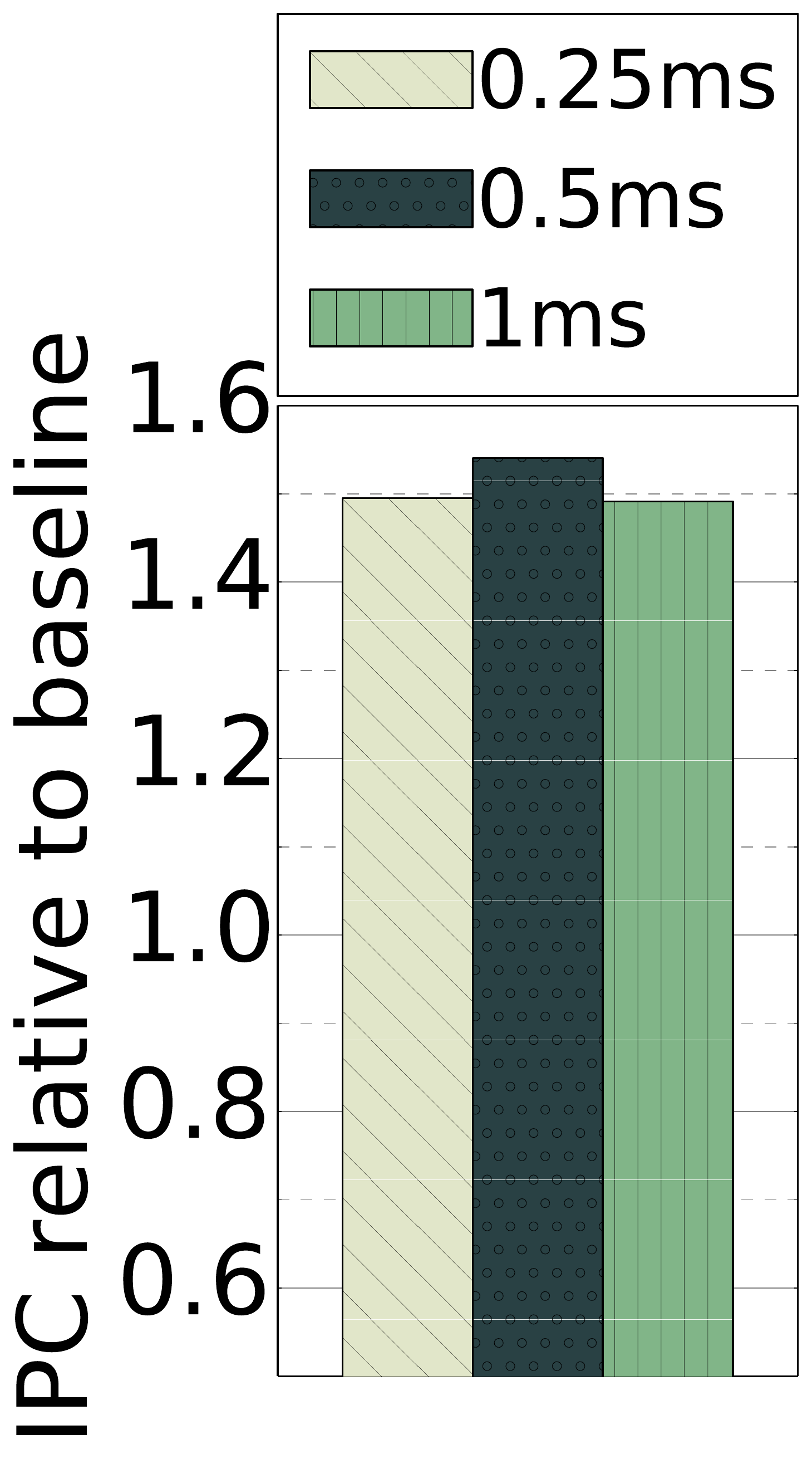}
\caption{Sensitivity for sample period in prefetch throttling.}
\label{fig:sen_sample_period}
\end{subfigure}
\vspace{-0.5em}
\caption{Sensitivity analysis.}
\label{fig:sensitivity}
\end{figure}

The reconfiguration interval, determines how frequently the different resource allocation controllers are invoked when running a workload. 
We investigate the sensitivity of CBP to different reconfiguration interval values, in order to determine an appropriate interval. Figure \ref{fig:sen_recon} shows the average (geo. mean) performance when using three different reconfiguration intervals - 1ms, 10ms and 100ms. A shorter reconfiguration period has the potential to adapt faster to phase change behaviour. However, it also incurs a higher overhead because a larger fraction of the interval would be used up for IPC sampling (required for prefetch throttling). Overall, the results show that using a 10ms period provides a good trade-off between quick adaptation and the overhead incurred for IPC sampling.


\vspace{-0.5em}
\subsubsection{Impact of cache size}
\label{sec:size}
The results thus far assume that each tile has a baseline cache allocation of 512kb. The total LLC capacity assuming a 16-core tiled CMP would be 8MB. We next study the impact of changing the cache capacity available for a single tile to 1MB. Figure \ref{fig:sen_cache_size} shows the average performance achieved using CBP with different per-tile capacity normalized to the baseline configuration with the same capacity. The results show that increasing the available LLC capacity leads to a 5\% drop in aggregate performance with CBP. This performance drop can be attributed to an increase in cache access time for the larger cache.

\vspace{-0.5em}
\subsubsection{Impact of changing bandwidth partitioning parameters}
\label{sec:bw_sen}
We investigate the sensitivity to the minimum bandwidth allocation, used in the bandwidth allocation algorithm presented in Section \ref{sec:bw}. Figure \ref{fig:sen_min_bw} shows the difference in performance with a minimum bandwidth allocation of 0.5GB/s and 1GB/s, normalized to the baseline. There were small variations among the different workloads and some experience a slight improvement while others workloads experience a small drop in performance. Overall, reducing the minimum bandwidth allocation did not have a considerable impact on the performance of CBP as long as it is sufficient for workloads which experience very little queuing delay. 

\vspace{-0.5em}
\subsubsection{Impact of changing prefetch sampling interval}
\label{sec:pref_sen}
We finally investigate the impact of changing the  \textit{prefetch\_sampling\_period} used in the prefetch throttling controller \ref{sec:pref}. 
Figure \ref{fig:sen_sample_period} shows the impact of changing the sampling period on performance normalised to the baseline. The intervals we use for evaluation are 0.25ms, 0.5ms and 1ms. The advantage of using a shorter sampling period is that it carries a lower overhead, while the drawback is the risk of over/under estimating the performance benefit from prefetching. The results indicate the sampling interval of 0.5ms achieves the best performance.

\vspace{-1em}
\section{Related work}
\label{sec:relatedwork}
\textbf{Isolated Management:} Several techniques have been proposed in the literature that focus specifically on cache partitioning, bandwidth partitioning and prefetch throttling. Cache partitioning techniques ~\cite{Qureshi2006,Lee2011,kpart18,Manikantan2012,Sanchez2011} help improve performance and achieve better utilization of available cache resources, by avoiding interference among co-running applications and reducing the number of accesses to memory. 
Bandwidth partitioning techniques ~\cite{emba19,Park2018,pabst17,liu10}, reduce average memory access penalty, by dynamically determining how bandwidth must be shared among the co-running applications.
Prefetching can hide memory access latency by fetching the data before it is requested~\cite{dahlgren93,Dahlgren1996,Nesbit2004,Nesbit2005,Lee2012,Kondguli2018}. However, inaccurate prefetches can impact application performance since it can increase the number and cost of demand misses~\cite{Ebrahimi2009}. Prefetch throttling \cite{dahlgren93,FDP07}, involves adaptively tuning when and what prefetcher settings are used dynamically based on application characteristics and has been shown to provide better performance and address drawbacks of prefetching. 
The aforementioned works, consider each of the techniques in isolation and leaves room for improvement, as shown in this work, since they do not take the interaction between cache partitioning, bandwidth partitioning and prefetch throttling into account. 

\textbf{Coordinated Management:} Several works have proposed combining two of the techniques in order to exploit the benefits from coordination. These works can be broadly classified into the following groups: i) coordinated cache and bandwidth partitioning ~\cite{Sahu2014,copart19}, ii) coordinated prefetching and cache partitioning~\cite{Xiao2019,Sun2019}, and iii) coordinated bandwidth partitioning and prefetching~\cite{Liu2011}. Sahu et al. propose ~\cite{Sahu2014} a method for cache and bandwidth partitioning, using a CPI model for bandwidth and set partitioning for the cache. CoPart~\cite{copart19} combines bandwidth and cache partitioning using a user-level run-time. Unlike CBP, their goal is to improve fairness. Recently, CPpf~\cite{Xiao2019} and Sun et.al ~\cite{Sun2019} propose a coordinated approach for cache partitioning and prefetch where prefetch friendly applications where given a smaller cache allocation. 
Unlike CBP which maintains per-application partitions, cache partitioning in these two proposals is performed for groups of applications.
Ebrahimi et al. ~\cite{Ebrahimi2011} propose general mechanisms to make memory scheduling techniques prefetch aware. 
However, these works cannot use the additional interactions and trade-offs which are available when coordinately managing all three resources, which we have shown is important for performance.

Some works have also proposed coordinated management of multiple resources. For instance, 
CLITE ~\cite{clite20} uses bayesian optimization to provide theoretically-grounded resource partitioning to meet QoS targets of multiple resources (e.g., cores, caches, memory bandwidth, memory capacity, disk bandwidth etc.) among multiple co-located jobs. 
Bitirgen et al. ~\cite{Bitirgen2008} use machine learning to manage power, cache and bandwidth in a coordinated way to anticipate system-level performance impact of allocation decisions. However, in neither of these works is prefetch throttling considered, which we have shown is important in order to realise the full potential of coordinated resource management.
To the best of our knowledge, CBP is the first coordinated resource manager for cache partitioning, bandwidth partitioning and prefetch throttling.

\vspace{-0.5em}
\section{Conclusions}
\label{sec:conclusion}
We present CBP, a mechanism for coordinated management of  cache partitioning, bandwidth partitioning and prefetch throttling. The design is motivated by our in-depth characterisation of the performance impact of cache, bandwidth and prefetch allocation and their interactions. CBP combines local resource allocation controllers with a coordination mechanism that dynamically manages and allocates the resources, in a way which considers both inter- and intra-application interactions. Our evaluation, on a tiled 16-core CMP, demonstrates that CBP improves performance by up to 86\% (geo. mean 50\%) compared to a system without partitioning and prefetching and by up to 36\% (geo. mean 11\%) over the state-of-the-art technique that manages cache partitioning and prefetching in a coordinated manner.


\bibliographystyle{ACM-Reference-Format}
\bibliography{bibliography}




\end{document}